\documentclass[%
 reprint,
showpacs,
amsmath,amssymb,
 aps,
]{revtex4-1}

\usepackage{fancyhdr}
\usepackage{graphicx}
\usepackage{dcolumn}
\usepackage{bm}

\begin{document}

\title{(Carrier Drift Modulation and the) Hyperbolic Time Crystals}%

\author{Evgenii E. Narimanov${}^1$ and Boris Shapiro${}^2$}
\affiliation{${}^1$School of Electrical and Computer Engineering and Birck Nanotechnology Center, Purdue University, West Lafayette, Indiana 47907, USA}
\affiliation{${}^2$Department of Physics, Technion - Israel Institute of Technology, Haifa 32000, Israel}
\date{\today}

\begin{abstract}
We introduce the Carrier Drift Modulation -- a new mechanism for creating temporal boundaries and enabling photonic time crystals. This approach opens a direct route to hyperbolic temporal metamaterials and, in particular, hyperbolic time crystals. We demonstrate that the very process responsible for time crystal formation can simultaneously compensate intrinsic material losses in the supporting medium — overcoming one of the central challenges in nanophotonics. The realization of truly lossless hyperbolic media, long considered as  one of the key challenges of nanophotonics, unlocks new possibilities for subwavelength light focusing, strong‑field physics, and novel regimes of light–matter interaction. Crucially, the proposed approach can be implemented using existing materials and readily available light sources, making it both practical and transformative.
\end{abstract}

\maketitle

\section{Introduction}

Ultra-fast temporal modulation of an optical material \cite{Goulielmakis2008} -- where its  electromagnetic properties are rapidly varied over time -- creates dynamic boundaries that light can interact with, much like spatial interfaces in conventional optics.\cite{Morgenthaler1958,Mendonca2002} When these temporal changes occur faster than the oscillation period of the incident light, they give rise to distinctive phenomena known as time reflections and time refractions, in which photons experience abrupt changes in frequency and propagation direction, while their momentum remains conserved.\cite{Morgenthaler1958,Mendonca2002,Relly2007} When such modulation is applied periodically, the system evolves into a photonic time 
crystal \cite{Zurita-Sanchez2012,Shaltout2016,Cervantes2009,Segev2018,VSMotiR1,VSMotiR2,Li2023,Feinberg2025}  -- a medium with a repeating structure in time that enables novel forms of light control, from amplification to frequency conversion and precise manipulation of photonic states \cite{Segev2018,VSMotiR1,VSMotiR2}.

However, achieving optical modulation that is both strong and fast is a significant challenge, as balancing these extremes often strains material limits and device performance.\cite{Khurgin-index} Yet, even relatively weak modulation can radically alter optical response when it induces a transition \cite{Zheludev1989} -- such as from the dielectric to the hyperbolic regime \cite{Science2012}. Crucially, such topological transitions occur only in anisotropic media \cite{Science2012}, as isotropic materials cannot  support hyperbolic dispersion \cite{PN}.

Here we introduce a new approach to ultra-fast optical modulation capable of inducing a topological transition from elliptical to hyperbolic regimes in a fundamentally isotropic material. The proposed approach of  Carrier Drift Modulation, shifts the velocity distribution of free charge carriers in the direction of the average electric field of an ultra-fast optical pump pulse. This free carrier drift creates a transient, pump-induced anisotropy, enabling an optically driven electromagnetic topological transition from an elliptic to a hyperbolic response.

Our approach not only offers a solution to the challenge of achieving strong modulation at ultra-fast timescales, but also opens the path to realizing a Hyperbolic Time Crystal -- a photonic time crystal that inherits the exotic properties of hyperbolic media. By inducing transient anisotropy in an otherwise isotropic material, Carrier Drift Modulation enables periodic temporal transitions into the hyperbolic regime, combining the dynamic control of time crystals with the high-wavenumber mode support and extreme light-matter interaction characteristics of hyperbolic materials.

With the proper choice of the temporal modulation period, the Hyperbolic Time Crystal can also support parametric amplification, when the energy from the modulation is transferred to the optical field. This gain mechanism offers a pathway to compensate for intrinsic material losses, long considered the fundamental limitation of hyperbolic media, with the potential to reach the Holy Grail of nanophotonics -- the lossless hyperbolic medium.

\section{Carrier Drift Modulation}

An intense optical pulse propagating in a conducting sample, rapidly accelerates the free charge carriers -- leading to a noticeable increase 
of their kinetic energy above the Fermi level. With the ever present band non-parabolicity inherent to all conducting materials, this immediately 
leads to a change of the free carrier response to a time-dependent probe field -- offering a practical approach to ultra-fast modulation of 
electromagnetic response. In a typical setting of this framework, the high optical pump intensity  that is necessary for  strong modulation,
is achieved by the optical time-compression of laser pulses -- naturally leading to the eventual pulse duration $\tau_M$ that is relatively short on the ``electronic'' time scale $\tau_0$ defined by the rate of the free carrier scattering, but well above the time of a single optical pump cycle 
$2 \pi/\omega_M$:
\begin{eqnarray}
\frac{2 \pi}{\omega_M} \ll \tau_M \ll \tau_0.
\end{eqnarray}
In general, an optical pump pulse leads to both the ``drift'' of the free carriers in the direction of the applied electric field, $\langle {\bf p}\rangle \equiv {\bf p}_M \propto {\bf E}_M$, and to the increase of their average energy $\Delta\langle \varepsilon_{\bf p}\rangle  \propto \langle { {\bf E} _M}^2 \rangle$. When  $\tau_M \gg 2 \pi/\omega_M$, multiple oscillation of the pump field rapidly average out the imposed carrier drift, ${\bf p}(t>\tau_M)\to 0$.

However, the situation is very different when using ultra-short pulses that last for only a few cycles -- all the way to the unipolar 
(sub-cycle) fields \cite{Gorelov2025} : 
\begin{eqnarray}
\tau_M \lesssim \frac{2 \pi}{\omega_M}  \ll \tau_0.
\label{eq:Tm}
\end{eqnarray}
In this case, the drift momentum of the free carrier distribution imposed by the entire pump pulse, $\langle {\bf p}\rangle  \neq 0$, immediately leading to an {\it anisotropic} electromagnetic response of the modulated medium.

The free carrier electromagnetic response strongly depends on the signal frequency $\omega$, 
\begin{eqnarray}
{\rm Re} \left[ \epsilon\left(\omega\right)  \right] \propto 1 - \omega_p^2 / \omega^2
\end{eqnarray}
and changes sign at the plasma frequency $\omega_p$ that is defined by the free carrier dynamics. 
Therefore, the anisotropy of the dielectric permittivity tensor imposed by the drift modulation, implies the necessary existence of the frequency  band where the modulation turn the originally isotropic response (dielectric when $\epsilon_\omega > 0$ and metallic when $\epsilon_\omega<0$) into hyperbolic (for opposite signs of the permittivity components in the directions that are parallel and perpendicular to the modulating electric field). For such signal wavelength, the drift modulation introduced in the present work, will not simply quantitatively change the system parameters but will instead induce a {\it  topological transition} \cite{Science2012} from elliptic to hyperbolic dispersion, that is known \cite{Science2012} to {\it qualitatively} change the electromagnetic response of the material.

The mechanism of the pump-induced anisotropy of the electromagnetic response of free carriers
originates from the combined effect of the drift of the electron distribution in the direction defined by
the average electric field of the pump pulse, and the energy band non-parabolicity. With the energy-dependent effective mass of the charge carriers, energy transferred from the pump to the free
electrons, modifies their electromagnetic response \cite{Khurgin-index}. 
A strongly  anisotropic velocity distribution of the free electrons that was induced by the pump, will therefore result in an anisotropic non-parabolicity correction to the electromagnetic response -- leading to the
aforementioned transition in the appropriate frequency window.
Furthermore, since the  transition occurs at a time scale much
faster than the electronic relaxation time $\tau_0$,  irreversible
processes cannot suppress it. 
 
A timed sequence of short intense pulses incident on a conducting material (such as e.g., a doped semiconductor or a transparent conducting oxide (TCO) film), with the 
electric field ${\bf E}_M\left({\bf r}, t\right)$ acting on the free
electrons in the sample, creates a ``hot electron'' distribution $f_M\left({\bf p},t\right)$, which satisfies  the kinetic equation \cite{Ziman}
\begin{eqnarray}
\frac{\partial f_M}{\partial t} + {\bf v_p}\cdot\nabla f_M +  e {\bf E}_M\left({\bf r}, t\right) \cdot \frac{\partial f_M}{\partial {\bf p}} & = & - \frac{f_M - f_0}{\tau_0}, \ \ \ 
\label{eq:KE1}
\end{eqnarray}
with the initial condition of the system being in the equilibrium at  $t \to - \infty$ : 
\begin{eqnarray}
f_M\left( {\bf p}, {\bf r}; t \to - \infty\right) & = & f_{\rm 0}\left( {\bf p} \right),
\label{eq:IC}
\end{eqnarray}
where ${\bf v_p} \equiv \partial\varepsilon_{\bf p}/\partial {\bf p}$ is the charge carrier group velocity, and $f_0\left( {\bf p} \right)$ is the equilibrium (Fermi-Dirac) distribution that  is normalized to the free electron concentration $n_0$
\begin{eqnarray}
\frac{2}{\left( 2 \pi \hbar\right)^3} \int d{\bf p} \, f_0\left( {\bf p} \right) & = &  n_0.
\end{eqnarray}

The gradient term ${\bf v_p}\cdot\nabla f_M$  in the kinetic equation (\ref{eq:KE1}) describes the spatial nonlocality
of the electromagnetic response of the charge carriers, as the coordinate displacement of a free electron over the period
of a single optical cycle, scales as $v_F/\omega_M$, where $v_F$ is the free electrons Fermi velocity.  When 
$v_F \ll c$, this length scale is much smaller than the wavelength $\lambda_M
\sim  c / \omega_M$, and the  spatial gradient term in the kinetic 
equation (\ref{eq:KE1}) can be neglected.

For the solution of Eqn.  (\ref{eq:KE1}) with the initial condition (\ref{eq:IC}) we therefore obtain
\begin{eqnarray}
f_M
& = &
\int_{-\infty}^t \frac{dt'}{\tau_0}  \,
f_0\left({\bf p} - {\bf p}_M\left({\bf r}, t\right) + {\bf p}_M\left({\bf r},t'\right)   \right)
\, e^{-\frac{t - t'}{\tau_0} }, \ \ \ \label{eq:fM}
\end{eqnarray}
where ${\bf p}_M\left({\bf r}, t\right)$ is the impulse  that was transferred to the 
free electrons from the modulation field by the time $t$:
\begin{eqnarray}
{\bf p}_M\left({\bf r}, t\right) & = & e \int_{-\infty}^t {\bf E}_M\left({\bf r}, t'\right) \, dt'.
\label{eq:pD}
\end{eqnarray}
At the times shorter than the electronic relaxation, the solution (\ref{eq:fM}) corresponds to 
the reversible dynamics of a collision-less plasma \cite{LL-PK} under the action of the pump 
\begin{eqnarray}
f_M\left({\bf p},{\bf r};  t\right) & = & f_0\left( {\bf p} - {\bf p}_M\left({\bf r},  t\right)\right),
\label{eq:f}
\end{eqnarray} 
with ${\bf p}_M$ corresponding to the average (drift) quasi-momentum of the free carriers.

When the modulation results from a single pump pulse (of the duration $\tau_M$), 
 at its tail end 
\begin{eqnarray}
f_M\left({\bf p},{\bf r};   \tau_M \lesssim t \ll \tau_0 \right) 
& \simeq &   f_0\left( {\bf p} - {\bf p}_0\left({\bf r}\right) \right), \ \ \ \label{eq:f1}
\end{eqnarray}
while in its wake  ($t > \tau_M$) 
\begin{eqnarray}
f_M\left({\bf p},{\bf r};   t > \tau_M \right) & \simeq &  \left(f_0\left( {\bf p} - {\bf p}_0\left({\bf r}\right)  \right) -f_0\left( {\bf p}\right) \right) e^{- t/\tau_0} 
\nonumber \\ & + &  
f_0\left( {\bf p} \right), \label{eq:f2}
 \end{eqnarray}
where
\begin{eqnarray}
{\bf p}_0\left({\bf r}\right)  &=  & e \int_{-\infty}^\infty {\bf E}_M\left({\bf r}, t'\right) \, dt'
\label{eq:p0}
\end{eqnarray}
is the total impulse transferred to a free electron from the pump pulse.

In particular, for the Gaussian pump pulse
\begin{eqnarray}
E_M\left(t\right) & = & E_0 \ e^{-  t^2 / \tau_M^2} \cos\left(\omega_M t \right),
\end{eqnarray} 
we find
\begin{eqnarray}
{ p}_0& = & eE_0 \tau_M\,  \sqrt{\pi} \, e^{- \omega_M^2 \tau_M^2 / {4} },
\label{eq:p0g}
\end{eqnarray}
controlled by the product $\omega_M \tau_M / \pi$, which defines the number of significant cycles in the rapidly decaying pulse.
For an ultra-short pump pulse ($\omega_M \tau_M <  \pi$), the impulse  $p_0 \simeq e E_0 \tau_M$  for a doped semiconductor under typical experimental conditions can become of the oder of the Fermi momentum $p_F$. On the other hand, 
for $\omega_M \tau_M >  \pi$,  due to the cancellation between positive and negative regions 
of $E_M\left(t\right)$ the drift momentum $p_0$ becomes exponentially small -- so that  the pump pulse takes
 the free electrons to a strongly non-equilibrium state but then, reversibly, brings them back close to the equilibrium. It was the latter regime that was recently investigated in the experiments of Ref. \cite{Moti-Vlad}, where the pump pulse had roughly three significant cycles.

The Carrier Drift Modulation approach, that we introduced in the present work, corresponds to the case $\omega_M \tau_M <  \pi$, 
when the pump, almost instantaneously on free carrier time  scale, creates a new, anisotropic medium with drifting electrons. 

Prior to the action of the pump in this setting, the probe wave is a superposition of the conventional transverse plasmons with the dispersion relation 
\begin{eqnarray}
\omega = \sqrt{\omega_p^2 + k^2 c^2 / \epsilon_\infty}, \label{eq:w-initial} 
\end{eqnarray}
where $\epsilon_\infty$ is the lattice contribution to the dielectric permittivity and $k$ is the wavenumber. At the moment of the arrival of an abrupt Carrier Drift Modulation, the dispersion relation of the probe waves evolves into
\begin{eqnarray}
\omega' & = &  \sqrt{\omega_p^2 - \Omega^2 + k^2 c^2 / \epsilon_\infty},\label{eq:wprime}
\end{eqnarray}
where the correction $\Omega$ depends on the acquired drift momentum $p_0$ and the free carrier dispersion non-parabolicity. However, in order to satisfy the boundary conditions at the tempoerary interface induced by the arrival of the modulation pulse, the probe field
must also acquire components propagating in the directions opposite to those of the original wave. Both of these time-refracted and
time-reflected components now oscillate at the new freqeuncy $\omega'$, with the time-refraction and time-reflection amplitudes that scale as 
\begin{eqnarray}
{\cal T} \simeq \frac{\omega + \omega'}{2 \omega}, \ \ \ \  {\cal R} \simeq \frac{\omega - \omega'}{2 \omega} 
\end{eqnarray}
in the limit of small loss $\omega \tau_0 \gg 1$. The detailed analysis of the time-reflection and time-refraction in a drift-modulated medium
 will be presented in section \ref{sec:dfpm}.

Most importantly, the proposed Carrier Drift Modulation opens the route to the Hyperbolic Time Crystal, 
where the optical modulation not only turns an isotropic semiconductor into a hyperbolic medium, but at the same time also reduces its absorption via the mechanism of parametric amplification.

\section{The Wave Equation for Carrier Drift Modulation \label{sec:we}}

Here, we consider the electromagnetic response of an isotropic conducting medium under the Carrier Drift Modulation, introduced in the previous section.  For a probe field ${\bf E}\left({\bf r}, t\right)$  that is much smaller than the peak amplitude of the pump pulse $E_M$,
the resulting linear response can be described by the linearized kinetic equation \cite{Ziman}
\begin{eqnarray}
\frac{\partial g}{\partial t} + {\bf v_p}\cdot\nabla g +  e {\bf E} \cdot \frac{\partial f_M}{\partial {\bf p}} 
+  e {\bf E}_M \cdot \frac{\partial g}{\partial {\bf p}}
& = & 
- \frac{g}{\tau_0},
\label{eq:g}
\end{eqnarray}
where the complete distribution function
\begin{eqnarray}
f\left({\bf p}, {\bf r}; t\right) & = & f_M\left({\bf p}; t\right)  + g\left({\bf p}, {\bf r}; t\right).
\label{eq:fg}
\end{eqnarray}
With the  free carriers Fermi velocity $v_F \ll c$,  
the effects of spatial dispersion that originate from the second term at the right-hand side of Eqn. (\ref{eq:g}) can be neglected,  and we  obtain
\begin{eqnarray}
g
 = 
-  \int_{-\infty}^t & {dt'} &    e {\bf E}\left({\bf r}, t'\right) \cdot 
\frac{f_M\left({\bf p} - {\bf p}_M\left({\bf r}, t\right) + {\bf p}_M\left({\bf r},t'\right); t'   \right)}{\partial {\bf p}}
\nonumber \\
& \times & 
\exp\left( {-\frac{t - t'}{\tau_0} }\right). \ \ \ \label{eq:g1}
\end{eqnarray}
Substituting (\ref{eq:fM}) into (\ref{eq:g1}), we obtain
\begin{eqnarray}
g
&  = & 
- \frac{e^{- t/\tau_0} }{\tau_0}   \int_{-\infty}^t  {dt'}   
\int_{-\infty}^{t'} {dt''}  \,  e^{ t''/\tau_0} \nonumber \\
& \times & e {\bf E}\left({\bf r}, t'\right) \cdot 
\frac{\partial f_0\left({\bf p} - {\bf p}_M\left({\bf r}, t\right)  + {\bf p}_M\left({\bf r},t''\right)   \right)}{\partial {\bf p}},  \label{eq:g2}
\end{eqnarray}

Substituting into Maxwell's Equations the free current density defined by the distribution function $g\left({\bf p}, {\bf r}; t\right)$ from Eqn. (\ref{eq:g}), we obtain the Wave Equation (see Appendix \ref{sec:A-WE})
\begin{eqnarray}
& & \left(\frac{\partial}{\partial t}    +    \frac{1}{\tau_0} \right) \left(   \frac{\partial^2 {\bf E}}{\partial t^2} + 
\frac{c^2}{\epsilon_\infty}  \, {\rm curl}\, {\rm curl} \, {\bf E}  \right)
+  \omega_p^2 \ \frac{\partial{\bf E} }{\partial t} \nonumber \\
& = &  \frac{\partial }{\partial t}    \left[  {\bf M}_0 \cdot {\bf E}
+  e^{ -\frac{t}{\tau_0}} \int_{-\infty}^t  {dt'}  \, \, \frac{\partial {\bf M}}{\partial t}  
\cdot  {\bf E}\left(t'\right) \right], \ \ \ \ \ \ \ \label{eq:we2}
\end{eqnarray}
where the plasma frequency
\begin{eqnarray}
\omega_p^2 & = & \frac{4\pi n_0 e^2}{m_* \epsilon_\infty},
\end{eqnarray}
 the modulation kernel
 \begin{eqnarray}
 {\bf M}\left( t, t'\right) & \equiv & \omega_p^2  \int_{-\infty}^{t'} \frac{dt''}{\tau_0}  \, \exp\left({ \frac{ t''}{\tau_0}}\right) 
\nonumber \\
& \times & \left( 1  - \frac{m_*}{{\bf m}\left[ {\bf p}_M\left(t\right)  - {\bf p}_M\left(t''\right)  \right]}  \right), 
\label{eq:MM}
 \end{eqnarray} 
 and
\begin{eqnarray}
{\bf M}_0\left( t \right) \equiv {\bf M}\left( t, t \right) \,  e^{ -\frac{t}{\tau_0}}.  
\label{eq:M0}
\end{eqnarray}
Here, the modulated  effective mass tensor
\begin{eqnarray}
\frac{1}{m_{\alpha\beta}\left[{\bf q}\right]} & = & 
\biggl< \frac{\partial^2 \varepsilon_{\bf p+ q}}{\partial p_\alpha \partial p_\beta}  \biggr>_{\bf p}  
\end{eqnarray}
with the phase space average defined as 
\begin{eqnarray}
  \bigl< \,  F\left[{\bf p +  q}\right] \, \bigr>_{\bf p}  
  & \equiv & 
  \frac{\int d{\bf p} \, F\left({\bf p + q}\right) f_0\left({\bf p}\right)}{\int d{\bf p} \,  f_0\left({\bf p}\right)}.
 \end{eqnarray} 
For a parabolic band 
\begin{eqnarray}
  \biggl< 
  \frac{\partial^2  \varepsilon_{{\bf p+q}} }{\partial p_\alpha \partial p_\beta} \biggr>_{\bf p}  & = & 
  \biggl< \frac{\partial^2  \varepsilon_{{\bf p}} }{\partial p_\alpha \partial p_\beta}  \biggr>_{\bf p} \equiv 
  \frac{\delta_{\alpha\beta}}{m_*}
\end{eqnarray}
is defined by the average effective mass $m_*$, so that 
${\bf M} = 0$, the wave equation does not depend on the pump, and  the modulation has no effect on the free
carrier response to the probe field.

 However, substantial  electronic non-parabolicity  is inherent to the conducting materials used in electro-optics,  such as transparent conducting oxides \cite{TCOs,ITO} and semiconductors \cite{nmat}. Furthermore, this is 
generally not a small effect \cite{Julia,Liu2014}: e.g., over the range of relevant quasi-momenta the effective mass varies 
by $\sim 30$\% in gallium arsenide \cite{Balkaitski1968,Cardona1961,Piller1966}
 and by a factor of $5$ in indium arsenide  \cite{Balkaitski1968,Shulman1965}. 

The general wave equation (\ref{eq:we2}) can be further simplified when the duration of the modulation pulse is smaller
than the time scale of a single electromagnetic cycle of the signal ${\bf E}\left({\bf r},t\right)$,
which reduces (\ref{eq:we2}) to 
\begin{eqnarray}
 \left(\frac{\partial}{\partial t}  \right.  & + &  \left.   \frac{1}{\tau_0} \right) \left(   \frac{\partial^2 {\bf E}}{\partial t^2} + 
\frac{c^2}{\epsilon_\infty}  \, {\rm curl}\, {\rm curl} \, {\bf E}  \right)
+  \omega_p^2 \ \frac{\partial{\bf E} }{\partial t} \nonumber \\
& = & {\bf M}_0 \cdot \left( \frac{\partial }{\partial t} - \frac{1}{\tau_0} \right) \,{\bf E}. \ \ \ \ \ \ \ \label{eq:we3}
\end{eqnarray}
In this approach (see Appendix \ref{sec:A-BC}), the effect of the drift modulation leading to an abrupt change of ${\bf M}\left(t,t '\right)$, is described by
{\it three} independent  
boundary conditions at the time of the ``temporal boundary'' induced by the pump pulse. These boundary conditions correspond 
to the continuity of the fields

(i) ${\bf E}$, 

(ii) ${\bf F}$, and 

(iii) $\partial{\bf F}/\partial t$, 

\noindent
where the temporal displacement field ${\bf F}$ is defined as 
\begin{eqnarray}
{\bf F} & \equiv & \frac{\partial {\bf E}}{\partial t} - e^{{ -\frac{t}{\tau_0}}}
\int_{-\infty}^t  {dt'}  \, \, {\bf M}\left(t, t'\right)  
\cdot  {\bf E}\left({\bf r}, t'\right). \ \ \ \label{eq:F1}
\end{eqnarray}

When the electronic relaxation time $\tau_0$ is mush larger than other   characteristic time scale 
of the system dynamics (such as e.g.,  the modulation pulse duration $\tau_M$, the signal 
period $2 \pi/\omega$, and the time interval $T_M$ between different modulation pulses), the 
drift modulation wave equation can be reduced to (see Appendix \ref{sec:A-LL-we})
\begin{eqnarray}
  \frac{\partial^2 {\bf E}}{\partial t^2} & + &
\frac{c^2}{\epsilon_\infty}  \, {\rm curl}\, {\rm curl} \, {\bf E}  
+   \frac{m_* \,  \omega_p^2 }{{\bf m}_M\left(t\right) } \ {\bf E}  \nonumber \\
& = &  \omega_p^2  \ m_* \, \frac{d {\bf m}^{-1} }{dt} \int _{-\infty}^t dt'  \,{\bf E}\left(t'\right) .
\label{eq:weLL}
\end{eqnarray}
where the time-dependent effective mass tensor
\begin{eqnarray}
\frac{1}{ {\bf m}_M} & \equiv & \frac{1}{{\bf m}\left[ {\bf p}_M\left( t\right)  \right]}.
\label{eq:mt}
\end{eqnarray}
Furthermore, when the duration of the modulation pulse is smaller
than the time scale of a single electromagnetic cycle of the signal, Eqn. (\ref{eq:weLL}) reduces to
\begin{eqnarray}
 \frac{\partial^2 {\bf E}}{\partial t^2} & + &
\frac{c^2}{\epsilon_\infty}  \, {\rm curl}\, {\rm curl} \, {\bf E}  
+    \omega_p^2   \ \frac{m_* }{{\bf m}_M} \,  {{\bf E} }  = 0 .
\label{eq:weLL0}
\end{eqnarray}
with the continuity of the electric field ${\bf E}\left({\bf r},t\right)$ and the temporal displacement
\begin{eqnarray}
{\bf F} & = &  \frac{\partial {\bf E}}{\partial t} +  \, \omega_p^2   \ \frac{m_* }{{\bf m}_M}
\cdot \int_{-\infty}^t  {dt'}  \,   
 {\bf E}\left( t'\right). 
\label{eq:bcLL}
\end{eqnarray}

The general wave equation derived in the present section, that is applicable to general modulation field
${\bf E}_M\left({\rm r}, t\right)$  in an
arbitrary geometry, is one of the main results of the present work.

\section{The Dielectric Permittivity under Carrier Drift Modulation \label{sec:dpcdm}}

For a monochromatic probe field 
\begin{eqnarray}
{\bf E}\left({\bf r}, t\right) & = & {\bf E}_\omega\left({\bf r}\right)  \exp\left(  - i\omega t\right),
\label{eq:probe}
\end{eqnarray}
the free carriers polarization
\begin{eqnarray}
 & {\bf P}& \left({\bf r}, t\right)   =  \int_{-\infty}^t dt'\, {\bf j}\left({\bf r}, t'\right) \nonumber \\
& = & - \frac{\epsilon_\infty}{4 \pi}  \left[  \frac{\omega_p^2}{\omega \left(\omega + \frac{i}{\tau_0}  \right)  } 
+  e^{{ -\frac{t}{\tau_0}}}  \int_{-\infty}^{t}  {dt'}  \, e^{  \left(i \omega +  \frac{1}{\tau_0} \right) \left( t - t'\right)}    \right. 
\nonumber \\
& \times & \left. 
  \int_{-\infty}^{t'}  {dt''}  \, e^{  i \omega  ( t' - t'' )}  \,  
 {\bf M}\left(t', t''\right)    \right] {\bf E}\left({\bf r}, t\right)
 \label{eq:P1}
\end{eqnarray} 
The main contributions to the time integrals in (\ref{eq:P1}) come from the intervals $0 < t - t' < 1/\omega$ and  
$0 < t' - t'' < 1/\omega$.  If, and only if, the time that elapsed from the most recent {\it past} temporal boundary,
is much larger than the oscillation period at the probe frequency, $ 2 \pi/\omega$, the free carrier polarization can 
be expressed as
\begin{eqnarray}
& {\bf P}& \left( t\right)   =  - \frac{\epsilon_\infty}{4 \pi}  \left[  \frac{\omega_p^2}{\omega \left(\omega + \frac{i}{\tau_0}  \right)  } 
-   \frac{{\bf M}_0\left(t\right) }{\omega \left( \omega - \frac{i}{\tau_0}  \right) } 
     \right] {\bf E}\left(t\right), \ \ \ \ 
 \label{eq:P2}
\end{eqnarray}
with
\begin{eqnarray}
{\bf M}_0\left(t\right)  & = & \omega_p^2 \sum_{t_n < t} e^{- \frac{t - t_n}{\tau_0}} 
\left( \frac{m_*}{{\bf m}\left[ \Delta{\bf p}_{n}  \right]}   - \frac{m_*}{{\bf m}\left[ \Delta{\bf p}_{n-1}  \right]}  \right), 
\ \ \ \ \ \  
\end{eqnarray}
where
\begin{eqnarray}
\Delta{\bf p}_n & \equiv {\bf p}_M\left( t\right)  - {\bf p}_M\left(\frac{t_n + t_{n-1}}{2}\right)
\end{eqnarray} 
is the difference between the drift momenta imposed by different modulation pulses.

We can therefore introduce the (time-dependent) dielectric permittivity tensor $\boldsymbol{\epsilon}$ such that
\begin{eqnarray}
{\bf D} \equiv \epsilon_\infty {\bf E} + 4 \pi {\bf P} = \boldsymbol{\epsilon} \cdot {\bf E},
\end{eqnarray}
where
\begin{eqnarray}
\epsilon_{\alpha\beta}\left(\omega; t\right) & = & \epsilon_\infty \left( 1 - \frac{\omega_p^2}{\omega\left( \omega + i/\tau_0 \right)}\right) \delta_{\alpha\beta} \nonumber \\
&+  &  \epsilon_\infty \,  \frac{\Omega^{\ 2}_{\alpha\beta}\left(t\right)    }{ \omega \left( \omega - i/\tau_0\right)},
\label{eq:eps} 
\end{eqnarray}
and
\begin{eqnarray} 
\Omega_{\alpha\beta}^{\ 2} & = & 
\omega_p^2 \sum_{t_n < t} e^{- \frac{t - t_n }{\tau_0}} 
\left( \delta_{\alpha\beta}  - {m_*} 
\biggl< 
  \frac{\partial^2  \varepsilon_{{\bf p+ \Delta p}_n} }{\partial p_\alpha \partial p_\beta} \biggr>_{\bf p}   
   \right). \ \ \ \ \ \ 
\label{eq:Omega}
\end{eqnarray}

In the limit of moderate carrier drift modulation ($\Delta p  < p_F$), we find (see Appendix \ref{sec:A})
\begin{eqnarray}
\Omega_\parallel^{\ 2}\left( t\right)   & \simeq & {3} \, \frac{\alpha_* \omega_p^2}{m_0} \, 
 \sum_{t_n < t}   \Delta p_n^2 \, e^{- \frac{t - t_n }{\tau_0}},  \\ 
\label{eq:Wprl}
\Omega_\perp^{\ 2}\left( t\right) & \simeq &  \frac{\alpha_* \omega_p^2}{m_0} \, 
 \sum_{t_n < t}   \Delta p_n^2 \, e^{- \frac{t - t_n }{\tau_0}}.,
\label{eq:Wprp}
\end{eqnarray}
where the Kane's non-parabolicity parameter  $\alpha_*$ is  on the order of the inverse bandgap energy $E_g$ \cite{Kane1957},
and the subscripts correspond to the directions parallel ($\parallel$) and perpendicular ($\perp$) to the {\it pump} field
${\bf E}_M$ modulating the material.

Therefore,  in the wake of a single modulation pulse applied at $t = 0$, when $\tau_M \ll t \ll \tau_0$,  the dielectric permittivity tensor of the drift-modulated isotropic medium takes the form
\begin{eqnarray}
{\epsilon_{\parallel,\perp}}& = & {\epsilon_\infty}  \left( 1 -  \frac{\omega_p^2 - \Omega_{\parallel,\perp}^{\ 2}}{\omega^2} \right), 
\label{eq:epspp}
\end{eqnarray}
and describes hyperbolic response in the frequency band 
\begin{eqnarray}
\sqrt{\omega_p^2  -\Omega_\parallel^2}  < \omega < \sqrt{\omega_p^2  -\Omega_\perp^2}.
\label{eq:h-band}
\end{eqnarray}

Note however, that under the conditions of the carrier drift modulation, the time-dependent dielectric permittivity
can only be introduced at the time that is at least several probe cycles after a temporary interface.  

\section{Temporal Reflection \label{sec:dfpm}}

In its conventional setup, the temporal reflection occurs when the electromagnetic properties of a medium (such as its permittivity ) change suddenly in time, while remaining uniform in space. This abrupt temporal discontinuity causes part of the wave to reflect -- not spatially, but temporally. Here we consider the 
temporal reflection that arises from the carrier drift modulation in an isotropic conducting material. 

\begin{figure}[htbp] 
   \centering
   \includegraphics[width=2.5in]{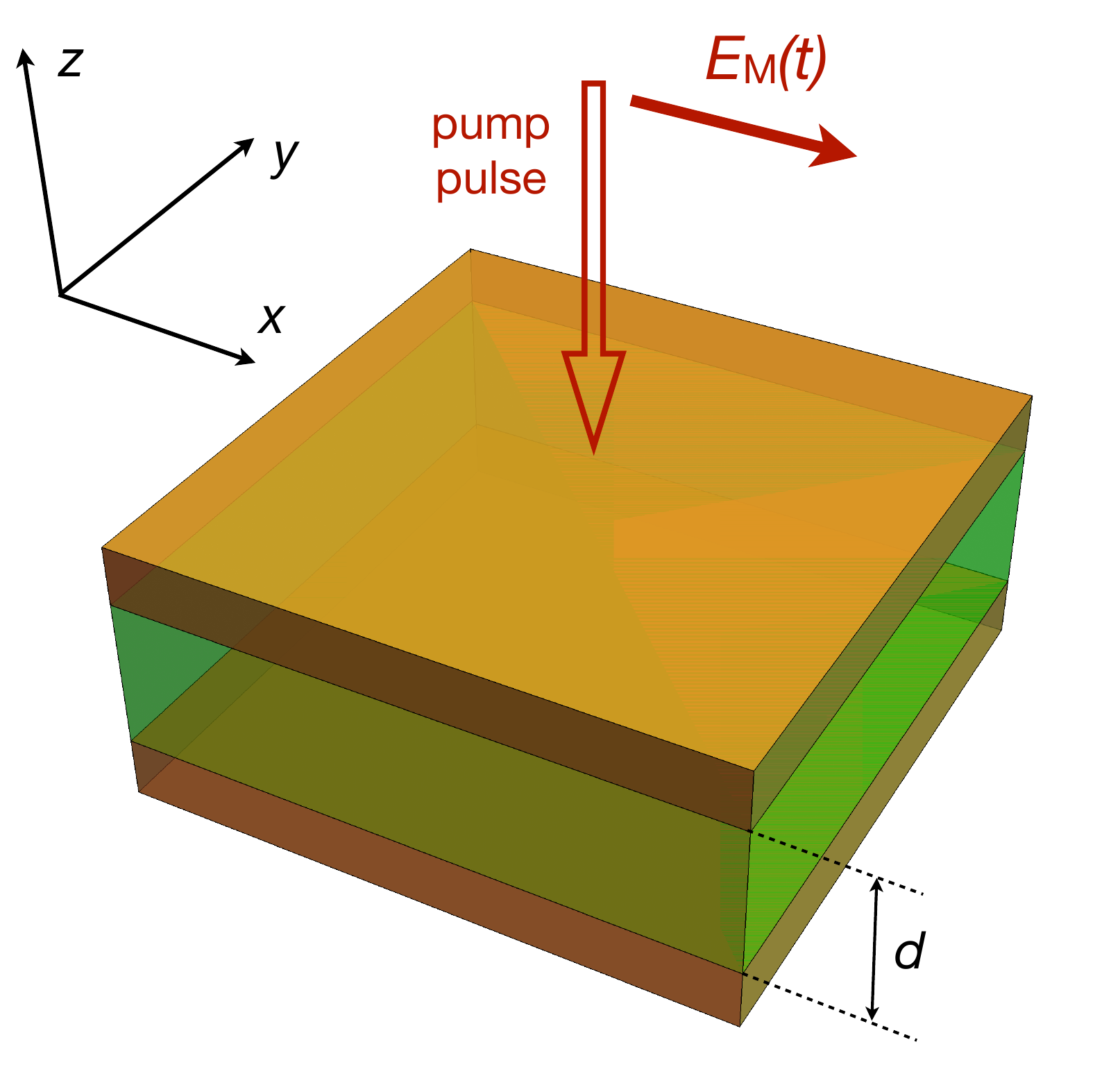} 
   \caption{Schematic of the waveguide geometry under drift modulation. The waveguide core consists of a conducting material with plasma frequency $\omega_p$, chosen to be comparable to the probe (signal) frequency $\omega$. The cladding behaves as a metal at the signal wavelength while remaining partially transparent to the high‑frequency optical pump pulse . The pump is polarized within the plane of the waveguide, and the signal propagates in the transverse electromagnetic (TEM) mode of the structure.}
   \label{fig:waveguide}
\end{figure}

To avoid the complexity of the retardation of the pump pulse as it propagates through the medium, we consider the geometry of a subwavelength waveguide with metallic cladding that is transparent to the pump 
(supported by the bandwidth that is well above the plasma frequency of the cladding) -- see Fig. \ref{fig:waveguide}. For a practical implementation, such system can be realized using a doped semiconductor waveguide (with the plasma frequency in mid- to far-infrared), with a thin conducting oxide cladding (such as e.g., indium tin oxide with the plasma wavelength close to  the telecom band around $1.55 \ \mu$m). 
 
We consider a guided wave, that was propagating at the frequency $\omega > \omega_p$ before the onset of the modulation ($t<0$). For the wave equation (\ref{eq:we3}) this corresponds to ${\bf M} = 0$, leading to
the ``incident'' wave
\begin{eqnarray}
{\bf E}_i\left({\bf r}, t\right) & = & {\bf E}_0\left({\bf r}\right)  \exp\left( - i\omega t\right),
\label{eq:Ei}
\end{eqnarray} 
in the geometry of Fig. \ref{fig:waveguide}, where 
\begin{eqnarray}
{\bf E}_0\left({\bf r}\right) & = & E_0 \, {\bf f}\left(k_z z\right) \, e^{i k_x x + i k_y y}
\end{eqnarray}
and
\begin{eqnarray}
\omega & = & k c/\sqrt{\epsilon_\omega},
\label{eq:wkc}
\end{eqnarray}
with the dielectric permittivity $\epsilon_\omega$  defined by Eqn. (\ref{eq:epspp}) with $\Omega_{\alpha\beta}$ set to zero. Note that due to the presence of the electromagnetic absorption that is inherent to conducting materials, 
at least one of the quantities $k$ or $\omega$ in Eqns. (\ref{eq:Ei}), (\ref{eq:wkc}) has a nonzero imaginary part.

When the waveguide thickness $d$ (see Fig. \ref{fig:waveguide}) is well below the wavelength at the signal frequency, the only propagating mode in the system is the so called TEM-wave \cite{Ramo-book} with the electric field that is perpendicular to the waveguide plane and only depends on the in-plane coordinates:
\begin{eqnarray}
{\bf f}_{\rm TEM} & = & \hat{\bf z},  \ \
k_z  =  0. 
\end{eqnarray}
For the ${\rm TEM}$-polarized field , when the pump pulse duration $\tau_M$ is much shorter than the modulation interval 
$T_M$ and the free carrier relaxation time $\tau_0$,  from Eqns. (\ref{eq:MM}),(\ref{eq:M0}) 
we find
\begin{eqnarray}
{\bf M}\left(t, t'\right) & = & {\bf \hat{z} \,  \hat{z}} \  \theta\left(t\right)  \exp\left(  \frac{{\rm min}\left[t', 0 \right]}{\tau_0} \right)  \, \Omega_\perp^2, \\
{\bf M}_0\left(t\right) & = & {\bf \hat{z} \,  \hat{z}} \  \theta\left(t\right)   \exp\left( -  \frac{t}{\tau_0} \right) \, \Omega_\perp^2, 
\end{eqnarray}
where  (see Fig. \ref{fig:waveguide})
\begin{eqnarray} 
\Omega_\perp^{\ 2} & = & 
\omega_p^2 
\left(1 - {m_*} 
\biggl< 
  \frac{\partial^2  \varepsilon_{{\bf p}+ {\hat{\bf x}} p_0} }{\partial p_z^2} \biggr>_{\bf p}   
   \right) \simeq  \frac{4}{3} \frac{\alpha_* p_0^2}{m_0} \, 
 \omega_p^2 , \ \ \ \ \ \ 
\label{eq:OmegaPP}
\end{eqnarray}
$\theta\left(t\right)$ is the Heaviside's function, and ${\bf p}_0$ was defined in Eqn. (\ref{eq:p0}). 

For the ${\rm TEM}$-polarized field at $0 < t \ll \tau_0$, we then obtain 
\begin{eqnarray}
{\bf E}  & = & \hat{\bf z} e^{ik_x x + i k_y y} \sum_{m=1}^3 s_m \exp\left( - i \omega_m t\right), 
\label{eq:Es}  
\end{eqnarray}
where the frequencies $\omega_m$ are the three solutions of the cubic equation
\begin{eqnarray}
\omega^3  +  \frac{i}{\tau_0} \omega^2 & - & \left( \omega_p^2 - \Omega_\perp^2  + \frac{k^2 c^2}{\epsilon_\infty} \right) \omega \nonumber \\ & = & \frac{i}{\tau_0} \left( \Omega_\perp^2  + \frac{k^2 c^2}{\epsilon_\infty} \right) ,
\end{eqnarray}
which can be equivalently expressed in the form
\begin{eqnarray}
\omega & = & \frac{kc}{\sqrt{\epsilon_\perp^{\rm eff}\left(\omega\right)}},
\label{eq:nu}
\end{eqnarray}
where the effective transverse permittivity
\begin{eqnarray}
\epsilon_\perp^{\rm eff}\left(\omega\right) & \equiv & \epsilon_\infty \left(1 - \frac{\omega_p^2 - \Omega_\perp^2 \left( 1 - \frac{i}{\omega \tau_0} \right)}{\omega \left(\omega + \frac{i}{\tau_0}\right)} \right) 
\end{eqnarray}
is consistent with $\epsilon_\perp$ in Eqn. (\ref{eq:epspp}). 

For a relatively small loss, when $c \, \tau_0 \gg 1/k $ \cite{explain} , we find
\begin{eqnarray}
\omega_{1,2} & = &   \pm \,  \omega' - i \, \frac{\gamma}{\tau_0}, \\
\omega_3 & = &  - i \, \frac{1 - 2 \gamma}{\tau_0}
\end{eqnarray}
where the new frequency 
\begin{eqnarray}
\omega' & = & \sqrt{\omega_p^2 - \Omega_\perp^2 +  {k^2 c^2}/{\epsilon_\infty}}, 
\label{eq:w-primed}
\end{eqnarray}
and the dimensionless extinction rate
\begin{eqnarray}
\gamma & = & \frac{1}{2} \, \frac{\omega_p^2 - 2 \Omega_\perp^2}{\omega_p^2 - \Omega_\perp^2  + {k^2 c^2}/{\epsilon_\infty}}. 
\end{eqnarray}
Therefore, in the electric field (\ref{eq:Es}) the amplitudes $s_1$ and $s_2$ correspond to the time-transmission and time-reflection coefficients  ${\cal R}$ and ${\cal T}$, while $s_3$ accounts for the amplitude of the time-evanescent wave ${\cal S}$ -- which yields 
\begin{eqnarray}
{\bf E} \left({\bf r}, t > 0\right)  & = & E_0  \, {\bf f}_{\rm TEM} \left\{ e^{- \frac{\gamma}{\tau_0} t 
+ i {\bf k}\cdot{\boldsymbol \rho} }  \left[ 
{\cal T} \,
e^{ - i \omega' t}  +
{\cal R} \,
e^{ i  \omega' t} 
\right] \right. \nonumber \\
& + & \left. 
e^{- \frac{1 - 2 \gamma}{\tau_0} t + i {\bf k}\cdot{\boldsymbol \rho}} 
\,
 {\cal S} \right\}, 
\label{eq:Es2}
\end{eqnarray} 
where $\boldsymbol\rho \equiv (x,y)$.

From the continuity of the fields ${\bf E}$, ${\bf F}$ and $\partial {\bf F}/\partial t$  at the temporal interface $t=0$, 
  we obtain (see Appendix \ref{sec:A-TRT}):
\begin{eqnarray}
{\cal R} & = & \frac{\eta_\omega \omega ( \omega - \omega' + i \frac{1 -  \gamma}{\tau_0} ) 
- i \frac{1 - 2 \gamma}{\tau_0} 
(\omega' - i \frac{ \gamma}{\tau_0} ) }{2 \, \omega' \, \left( \omega' - i\frac{1 - 3 \gamma}{\tau_0} \right) }, \\
{\cal T} & = & \frac{\eta_\omega \omega ( \omega + \omega' + i \frac{1 -  \gamma}{\tau_0} ) 
+i \frac{1 - 2 \gamma}{\tau_0} 
(\omega' + i \frac{ \gamma}{\tau_0} ) }{2 \, \omega' \, \left( \omega' + i\frac{1 - 3 \gamma}{\tau_0} \right) },
 \\
{\cal S} & = & \frac{ {\omega'}^2  - \eta_\omega \omega \left( \omega + 2 i  \gamma/\tau_0\right)  + \gamma^2 / \tau_0^2}{{\omega'} ^2+{\left(1 - 3 \gamma\right)^2}/{\tau_0^2} },
\end{eqnarray}
In the low-loss limit $\omega\tau_0 \gg 1$ we find
\begin{eqnarray}
\omega' \to \sqrt{\omega^2 - \Omega_\perp^2} \to \sqrt{\eta_\omega} \, \omega,
\end{eqnarray}
so that 
\begin{eqnarray}
{\cal R} & \simeq & \frac{\eta_\omega \, \omega  (\omega - \omega'  ) }{{2 \, \omega'}^2 }
\to \frac{\omega - \omega'}{2 \, \omega}, \label{eq:Rll}  \\
{\cal T} & \simeq & \frac{ {\eta_\omega \,  \omega \,   (\omega + \omega'  ) }}{{2 \, \omega'}^2 }
\to \frac{\omega + \omega'}{2 \, \omega} , \label{eq:Tll} \\
{\cal S} & \simeq & \frac{ {\omega'}^2  - \eta_\omega \, \omega ^2  }{{\omega'} ^2} \to 
 0. \label{eq:S0} 
\end{eqnarray}

Note that the conventional approach \cite{Morgenthaler1958,Mendonca2002}  based on the idea 
of a nearly-instantaneous change in the dielectric permittivity tensor in both time and frequency, can not reproduce the results of the present section, even in the lossless limit when 
the presence of the time-evanescent wave beyond the temporal interface can be neglected (see Eqn. (\ref{eq:S0}) ). Since the drift modulation rapidly changes the {\it current density}, the resulting dielectric polarization ${\bf P} \propto \int dt \, {\bf j}\left(t\right)$ remains {\it continuous} across the transition --  which cannot be described by a  step-like
behavior of the time-dependent dielectric permittivity \cite{Morgenthaler1958,Mendonca2002}.

\begin{figure}[htbp] 
   \centering
   \includegraphics[width=3.in]{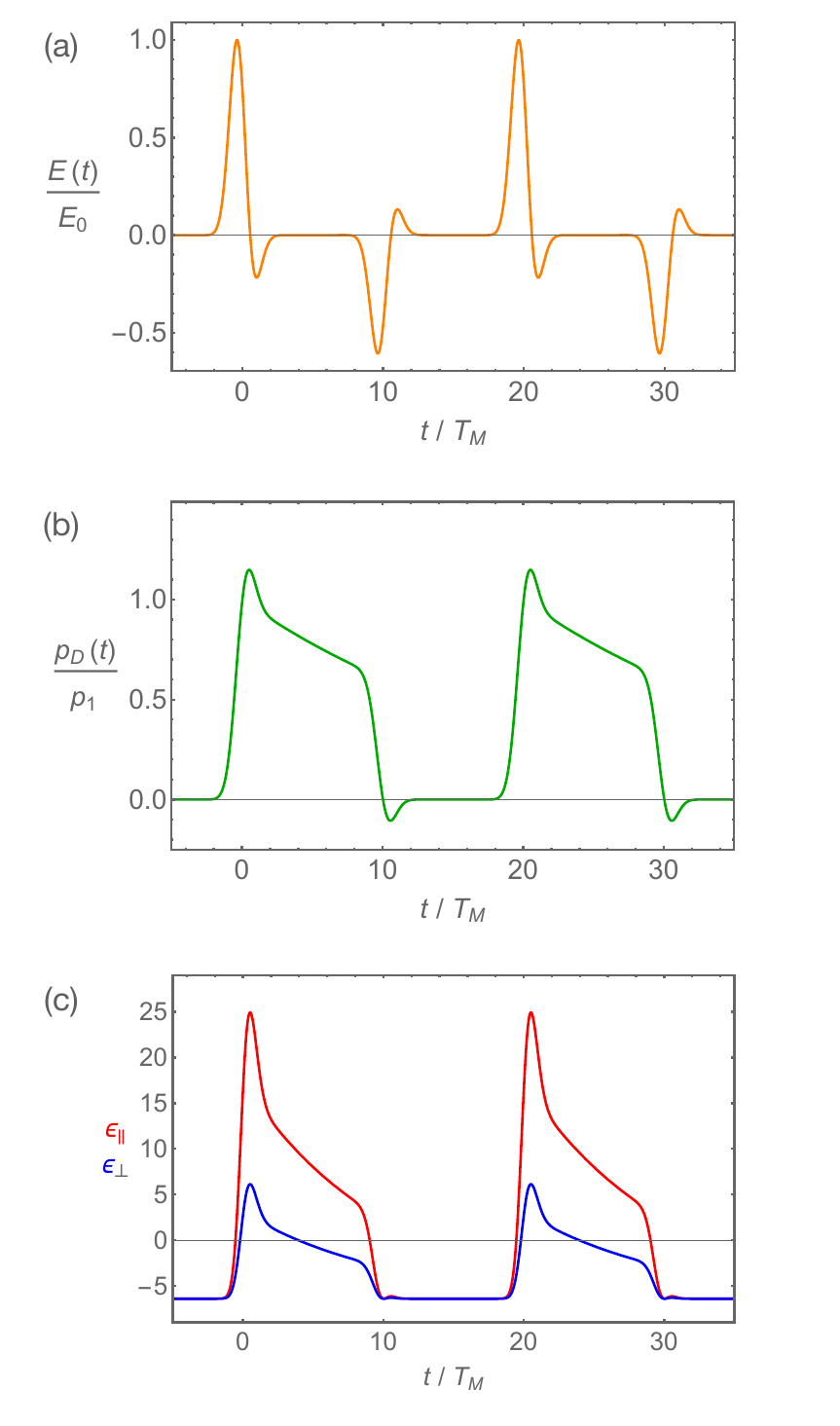} 
   \caption{Panel (a): Modulation pulse sequence used for hypercrystal drift modulation  (see Eqns. (\ref{eq:EM1}),(\ref{eq:EM})). Note the nonzero average field for a single modulation pulse, and alternating signs of the pulse train. Panel (b): Drift momentum of the free carriers, calculated as the average over the free electron distribution. Note that the carrier drift momentum $p_D$  is distinct from the pump impulse $p_M$. Panel (c): Time‑dependent dielectric permittivity of the medium, obtained from the electrical displacement vector ${\bf D}\left(t\right)$ corresponding to the time‑dependent distribution function of the free carriers $f\left({\bf p}, t\right)$. The ``parallel” ($\parallel$)  and ``perpendicular” ($\perp$) directions (shown by red and blue curves, respectively) are defined relative to the modulation field ${\bf E}_M$ (see also  Fig. \ref{fig:waveguide}). Numerical values used in the calculation correspond to a heavily doped gallium arsenide 
sample \cite{nmat}, with the relaxation time $\tau_0 = 0.1$ and the modulation interval $T_M = 20$~fs.
   }
   \label{fig:modulation}
\end{figure}

\section{The Hyperbolic Time Crystal}

With the distortion of the free electron velocity distribution
in the direction of the electric field of the pump pulse,  Carrier Drift Modulation creates a transient anisotropy  in the free carrier electromagnetic response -- see Eqn. (\ref{eq:epspp}). With the proper choice of the amplitude and phase for modulation pulse,  coherent pulse train can ``imprint'' the desired electron drift momentum  profile ${\bf p}\left( t\right)$, as long as the time interval
between the modulation pulses $T_M$ is substantially smaller than the electron scattering time $\tau_0$.  In particular, the ``binary phase-shift keying'' modulation sequence,
\begin{eqnarray}
{\bf E}_M\left(t\right) & = & \sum_{n} \left[ {\bf E}_1\left(t - 2 n T_M\right) \right. \nonumber \\
& - &   \left. e^{- T_M/\tau_0} \, {\bf E}_1\left(t - \left( 2 n-1\right) \,  T_M\right) \right],
\label{eq:EM1}
\end{eqnarray}
or, in the limit $\tau_0 \ll T_M$
\begin{eqnarray}
{\bf E}_M\left(t\right) & = & \sum_{n}  \left( -1\right)^n {\bf E}_1\left(t - n \, T_M\right),
\label{eq:EM}
\end{eqnarray}
where ${\bf E}_1\left(t\right)$ is a single ultra-fast optical pulse, will leads to the ``on-off'' drift momentum modulation -- 
see Fig. \ref{fig:modulation}. 

With the relatively low absorption in the original isotropic material at the probe/signal frequency $\omega$,  
\begin{eqnarray}
\omega \tau_0 \gg 1,
\end{eqnarray}
when the signal is injected into the system after it already reached the modulated steady state,
the photonic time crystal defined by the Carrier Drift Modulation (\ref{eq:EM}) operates as a 
 temporal metamaterial  when $ \omega \,  T_M \ll 1 $, and as the time hypercrystal for  $ \omega \,  T_M  \gtrsim 1$.
 
Signal propagation in the temporal metamaterial regime is describes by the standard wave equation with the effective dielectric permittivity tensor
 \begin{eqnarray}
 \epsilon_{\parallel,\perp} & = & \epsilon_\infty \left( 1 - \frac{\omega_p^2 - \langle \, \Omega^2_{\parallel,\perp}  \rangle}{\omega\left( \omega + i/\tau_0 \right)}\right). 
 \label{eq:eps-eff}
 \end{eqnarray}
where $\langle\ldots\rangle$ represents the time average over a full period of the modulation pulse train (e.g., $2 \,  T_M$ for
the binary phase-shift protocol of Eqn. (\ref{eq:EM})), with the hyperbolic response in the signal frequency band
\begin{eqnarray}
\sqrt{\omega_p^2  - \langle \, \Omega^2_{\parallel}  \rangle }  < \omega < 
\sqrt{\omega_p^2  - \langle \, \Omega^2_{\perp}  \rangle }.
\end{eqnarray}

The temporal hypercrystal regime  (defined by the requirement $ \omega \, T_M \gtrsim 1$) offers the temporal analogue of the ``spatial hypercrystals'' \cite{hypercrystal}, the composites with the unit cell that contains both hyperbolic media and natural materials. Similar to the hyperbolic material, the hypercrystal also supports propagating fields with the  wavenumbers that are unlimited by the signal frequency, albeit with a nontrivial bandstructure and multiple bandgaps.

However, with the proper choice of the signal frequency ($\omega  T_M  = \pi$), carrier drift modulation now satisfies the standard requirement for  the parametric amplification \cite{parametric_amplification} -- leading to the effective signal gain in the drift-modulated medium. With the general high-efficiency of the parametric amplification processes that turned the corresponding devices into the workhorses of high-bandwidth electronics \cite{parametric_electronics}, the parametric gain in the time hypercrystal can reach to the level of the signal absorption in the original (unmodulated) conducting medium -- which offers, for the first time, the possibility of a lossless hyperbolic medium.

At the length scales relevant for for high-wavenumber fields propagating in a hyperbolic medium, the spatial variation of the pump pulse can be neglected, so to avoid dealing with the pump retardation we are no longer limited to the waveguide geometry of the previous section. Furthermore, with the signal / probe primarily supported by the high wavenumbers $k \gg \omega/c$,
its propagation can be described within the framework of 
 the quasistatic limit \cite{LL-ECM}, where the electric field is defined by the scalar potential
\begin{eqnarray}
{\bf E} & = & - \nabla \phi.
\end{eqnarray}

For the wave equation in the quasistatic limit we  obtain (see Appendix \ref{sec:A-WE-QS}) 
\begin{eqnarray}
 \frac{d^2\phi_{\bf k} }{d t^2} & + & \frac{1}{\tau_0} \frac{d\phi_{\bf k} }{d t}  +   
 \left( \omega_p^2  - {\cal M}_{\bf k}^{(o)}\left(t\right) \right) \phi_{\bf k} 
 \nonumber \\
 & = & e^{ - \frac{ t}{\tau_0}} \,  \int_{-\infty}^t dt' \, \frac{\partial {\cal M}_{\bf k}}{\partial t}  \,  \phi_{\bf k}\left( t'\right), 
\label{eq:phi2}
\end{eqnarray}
where the (scalar) modulation kernel ${\cal M}_{\bf k}\left(t\right)$ is given by
\begin{eqnarray}
{\cal M}_{\bf k}\left(t, t'\right) & = & \frac{{\bf k}\cdot{\bf M}\left(t, t' \right)\cdot{\bf k}}{k^2}, 
\end{eqnarray}
and
\begin{eqnarray}
{\cal M}_{\bf k}^{(0)}\left(t\right)  & \equiv & e^{ - \frac{ t}{\tau_0}} \, {\cal M}_{\bf k}\left(t,t\right)  
\end{eqnarray}

When the duration of the modulation pulse is smaller
than the time scale of a single electromagnetic cycle of the signal, Eqn.  (\ref{eq:phi2}) can be reduced to 
\begin{eqnarray}
\frac{d^2\phi_{\bf k} }{d t^2} +  \frac{1}{\tau_0} \frac{d\phi_{\bf k} }{d t}  +   
 \left( \omega_p^2  - {\cal M}_{\bf k}^{(o)}\left(t\right) \right) \phi_{\bf k} 
 & = & 0 , 
\label{eq:phi3}
\end{eqnarray}
with the boundary conditions at the time of each modulation pulse corresponding to the continuity of
the (i) scalar potential $\phi$ and (ii) the auxiliary time displacement potential
\begin{eqnarray}
\psi\left({\bf r}, t\right)  \equiv \frac{\partial \phi}{\partial t} - e^{{ -\frac{t}{\tau_0}}}
\int_{-\infty}^t  {dt'}  \, {\cal M}_{\bf k}\left(t, t'\right)  
\cdot  \phi\left({\bf r}, t'\right). \ \ \ \label{eq:psi}
\end{eqnarray}

Neglecting the material absorption ($\omega \tau_0 \to \infty$), for the ``on-off" carrier drift modulation of Eqns. (\ref{eq:EM1}),(\ref{eq:EM}),
\begin{eqnarray}
{\cal M}_{\bf k}^{(0)}  & =  &
\left\{
\begin{array}{cc}
\xi_{\bf k} , & \lfloor t / T_M \rfloor = 2 n , \\
0,  &  \lfloor t / T_M \rfloor = 2 n + 1,
\end{array}
\right. \ \ \ \ \ 
\end{eqnarray} 
where $n$ is an integer, $\lfloor x \rfloor$ represents the integer part of $x$, and 
\begin{eqnarray}
\xi_{\bf k} & = &  \frac{k_\parallel^2}{k^2} \ \frac{\Omega_\parallel^2}{\omega_p^2} + \frac{k_\perp^2}{k^2} \  
 \frac{\Omega_\perp^2}{\omega_p^2}, 
 \label{eq:xi}
\end{eqnarray}
with the components ${\bf k}_\parallel$ and ${\bf k}_\perp$ parallel and perpendicular  to the 
modulation pulse field  ${\bf E}_1$.
Here 
\begin{eqnarray}
\Omega_\parallel^{\ 2}  & = &  \frac{\alpha_* p_1^2}{m_0}  \omega_p^2,  
\label{eq:WprlHC} \\ 
\Omega_\perp^{\ 2} &= & \frac{\alpha_* p_1^2}{m_0}  \omega_p^2,
\label{eq:WprpHC}
\end{eqnarray}
and $p_1$ is the transferred impulse from a single modulation pulse in (\ref{eq:EM1}), (\ref{eq:EM}):
\begin{eqnarray}
{\bf p}_1 & \equiv & \int_{-
\infty}^\infty dt \ e \, {\bf E}_1\left(t\right),
\end{eqnarray}

\begin{figure}[htbp] 
   \centering
   \includegraphics[width=3.25in]{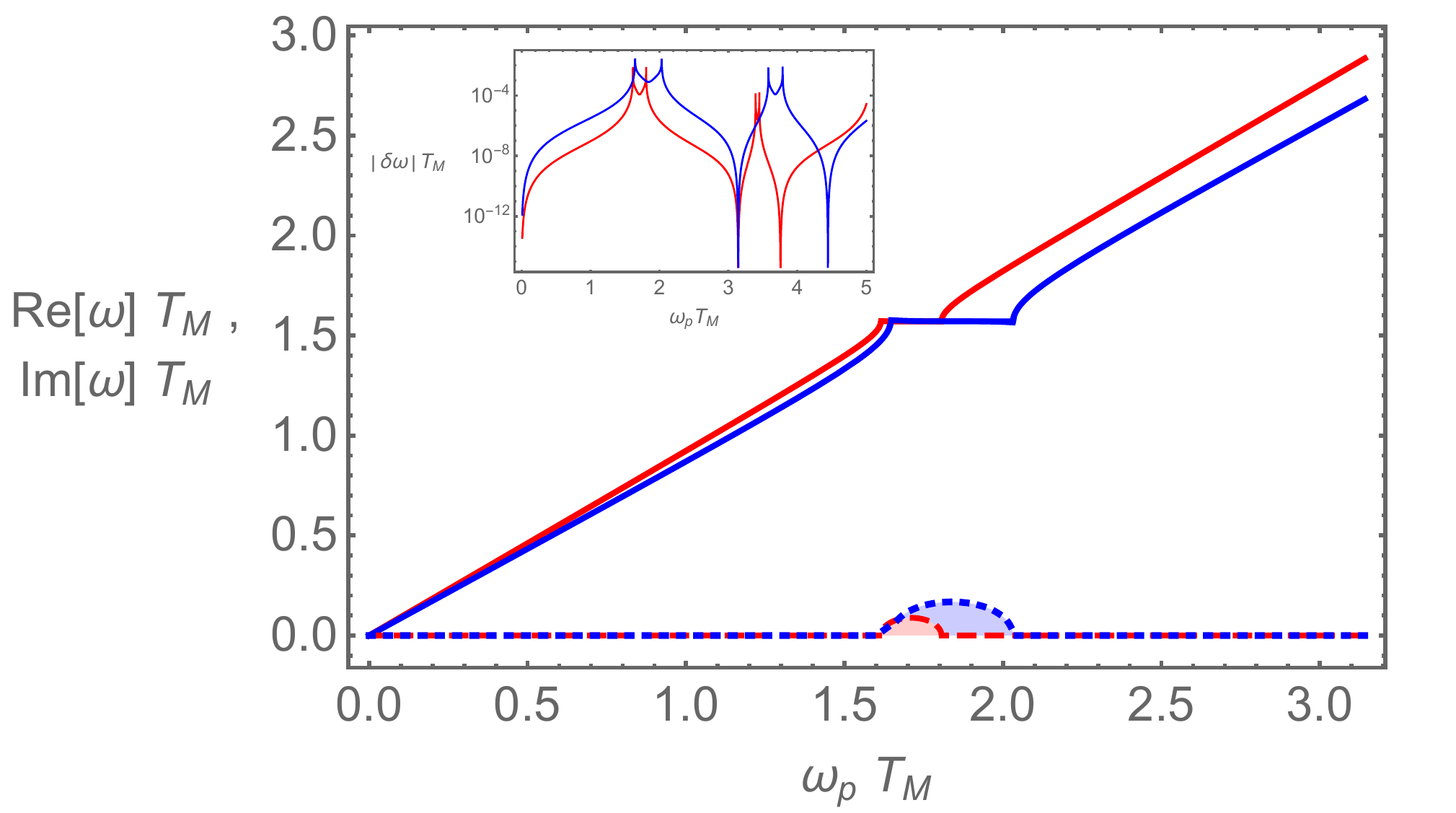} 
   \caption{Floquet–Bloch frequency in the drift‑modulated hypercrystal, shown as a function of the product of the material plasma frequency and the modulation interval for $\xi_{\bf k} = 0.3$ (red lines) and $\xi_{\bf k} = 0.5$ (blue curves). Solid and dashed lines represent the real and imaginary parts of the Floquet–Bloch frequency, respectively. The results of the exact solution (\ref{eq:det}) and its analytical approximation 
   (\ref{eq:wr}),  (\ref{eq:wi}) are indistinguishable in the main plot, with the small $(\lesssim 10^{-4}$)  difference between the exact and analytical solutions shown in the inset.}
   \label{fig:dispersion}
\end{figure}

For the Floquet-Bloch states in the resulting time crystal,ƒ
\begin{eqnarray}
{\bf E}\left({\bf r}, t\right) & = & {\bf u}\left(t\right) e^{i {\bf k}\cdot{\bf r} - i \omega t}, 
\end{eqnarray}
with
\begin{eqnarray}
{\bf u}\left( t\right)  =  {\bf u}\left( t + 2 \, T_M\right),
\end{eqnarray}
 we obtain
 \begin{eqnarray}
 & &   \cos  \left[  2 \, \omega    T_M \right]   =  
\cos\left[\omega_p T_M\right] \cos\left[\omega_{\bf k}  T_M \right]  
\nonumber \\
&  & - \frac{1}{2}   \left( \frac{\omega_{\bf k}}{\omega_p} +   \frac{\omega_p}{\omega_{\bf k}}  \right) 
\, 
  \sin\left[\omega_p T_M\right] \, \sin\left[\omega_{\bf k} T_M \right], 
  \label{eq:det}
 \end{eqnarray}
 where
 \begin{eqnarray}
 \omega_{\bf k} & \equiv \sqrt{\omega_p^2 -  \left( {k_\parallel}/{k} \right)^2 \,  \Omega^2_\parallel  
 -  \left( {k_\perp}/{k} \right)^2  \, \Omega_\perp^2 }.
 \end{eqnarray}
An accurate solution of Eqn. (\ref{eq:det}) can be expressed as
 \begin{eqnarray}
& &  {\rm Re} \left[ \omega \right]  =  \frac{\omega_p + \omega_{\bf k}}{2} 
  +  \frac{1}{T_M}  \tan\left[ \left( \omega_p + \omega_{\bf k} \right) T_M  \right]   
 \nonumber \\
 &  \times  & 
{\rm Re} \, \sqrt{  1 +
   \frac{\left(\sqrt{\frac{\omega_{\bf k}}{\omega_p} } - \sqrt{\frac{\omega_p}{\omega_{\bf k}} } \right)^2 \sin\left[\omega_p T_M\right] \, \sin\left[\omega_{\bf k} T_M \right]}{\sin\left[ \left( \omega_p + \omega_{\bf k} \right) T_M\right] \, \tan\left[ \left( \omega_p + \omega_{\bf k} \right) T_M\right] } }\nonumber \\
&-    &  \frac{1}{T_M} \,  \tan\left[ \left( \omega_p + \omega_{\bf k} \right) T_M  \right]  
  \label{eq:wr}
 \end{eqnarray}
and
 \begin{eqnarray}
& &  {\rm Im} \left[ \omega \right]  =  {\rm Im} \  \frac{1}{T_M} \, 
 \Bigg[ \tan^2\left[ \left( \omega_p + \omega_{\bf k} \right) T_M\right] 
 \nonumber \\
 &  + &  \left(\sqrt{\frac{\omega_{\bf k}}{\omega_p} } - \sqrt{\frac{\omega_p}{\omega_{\bf k}} } \right)^2
   \frac{ \sin\left[\omega_p T_M\right] \, \sin\left[\omega_{\bf k} T_M \right]}{\cos\left[ \left( \omega_p + \omega_{\bf k} \right) T_M\right]  }  \Bigg]^{1/2}. \ \ \ 
   \label{eq:wi}
 \end{eqnarray}
 In Fig. \ref{fig:dispersion} we show the exact solution of Eqn. (\ref{eq:det}), with absolute error of the analytical  expressions
 (\ref{eq:wr}), (\ref{eq:wi}) in its inset. 
 
In the time metamaterial regime $\omega_p T_M \ll 1$, Eqn. (\ref{eq:det}) reduces to
\begin{eqnarray}
 \omega^2 & = & \left( 1 - \frac{\xi_{\bf k}}{2}\right)   \omega_p^2.
\label{eq:tmm1}
\end{eqnarray} 
Substituting (\ref{eq:xi}) into (\ref{eq:tmm1}), we obtain the standard effective medium dispersion 
\begin{eqnarray}
\epsilon_\parallel k_\parallel^2 + \epsilon_\perp k_\perp^2 = 0,
\label{eq:tmm2}
\end{eqnarray}
where the effective permittivity tensor
\begin{eqnarray}
\epsilon_{\parallel,\perp} & = & \epsilon_\infty \left(1-\frac{ \omega_p^2 - \frac{1}{2}\,{\Omega_\parallel^2}}{\omega ^2} \right) 
\label{eq:eps-tmm}
\end{eqnarray}
is consistent with (\ref{eq:eps-eff}). As expected, in the frequency range (\ref{eq:h-band}) the time metamaterial behaves as a hyperbolic medium.

In contrast to this behavior, when $\omega_p T_M \sim \pi n /2$ for any integer $n$, the photonic time crystal formed in the
proposed drift modulation approach, shows bandgaps with ${\rm Im}\left[\omega\right] > 0$. This corresponds to the 
photonic time crystal gain \cite{VSMotiR1}, whose physical origin lies in the optical parametric amplification due to the periodic
optical modulation of the medium. 

\begin{figure}[htbp] 
   \centering
   \includegraphics[width=3.5 in]{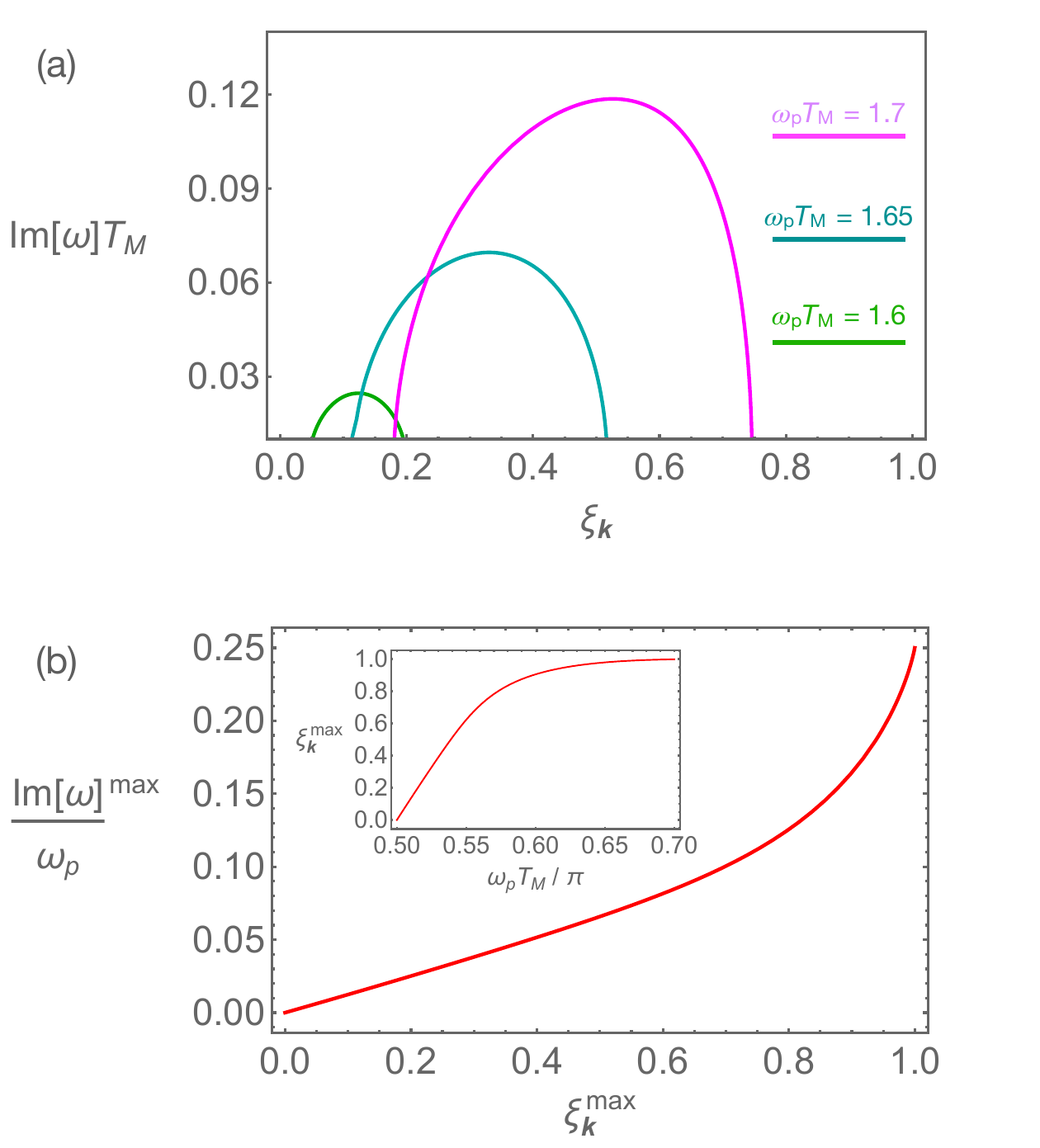} 
   \caption{Parametric gain in the photonic bandgap of the hyperbolic time crystal. Panel (a): Parametric gain as a function of
   $\xi_{\bf k}$ for different values of the product $\omega_p T_M$.Panel (b): Parametric gain as a function of
   $\xi_{\bf k}^{\rm max}$, where $\xi_{\bf k}^{\rm max}$ is defined as the value of $\xi_{\bf k}$ corresponding to the maximum gain for a given value of $\omega_p T_M$. The inset shows the dependence of $\xi_{\bf k}^{\rm max}$ on $\omega_p T_M / \pi$.}
   \label{fig:gain}
\end{figure}

For a given material with its plasma frequency $\omega_p$, the choice of the modulation
interval $T_M$ defines the interval of $\xi_{\bf k}$ with a nonzero gain -- see Fig. \ref{fig:gain}(a). Naturally, due to a finite absorption in the system, 
it will primarily support the waves with the largest gain ${\rm Rm}\left[ \omega\right]$, and thus the smallest amount of 
attenuation. In Fig. \ref{fig:gain}(b) we show how the corresponding value $\xi_{\bf k}^{\rm max}$ varies with $\omega_p T_M$
(inset) and plot the resulting maximum gain as a function of  $\xi_{\bf k}^{\rm max}$.

From Eqn. (\ref{eq:xi}), for a fixed value of $\xi_{\bf k}^{\rm max}$ we find
\begin{eqnarray}
\left( \xi_{\bf k}^{\rm max} - \frac{\Omega_\parallel^2}{\omega_p^2}\right) k_\parallel^2  
+\left( \xi_{\bf k}^{\rm max} - \frac{\Omega_\perp^2}{\omega_p^2}\right) k_\perp^2  & = & 0,
\end{eqnarray}
which implies hyperbolic dispersion when
\begin{eqnarray}
\frac{\Omega_\perp^2}{\omega_p^2} < \xi_{\bf k}^{\rm max} <  \frac{\Omega_\parallel^2}{\omega_p^2},
\end{eqnarray}
or (see Eqns. (\ref{eq:WprlHC}), (\ref{eq:WprpHC}) 
 \begin{eqnarray}
 \frac{\alpha_* p_1^2}{m_0} < \xi_{\bf k}^{\rm max} <   {3} \frac{\alpha_* p_1^2}{m_0}.
 \label{eq:h-band-p}
\end{eqnarray}
 With a continuous variation of $ \xi_{\bf k}^{\rm max}$ as function of the modulation interval $T_M$  (see the inset of Fig. \ref{fig:gain}(b)),
 the range (\ref{eq:h-band}) should be easily accessible in an experiment.
 
 Furthermore, in highly doped semiconductors such as GaAs and InGaAs \cite{nmat}, the ``quality factor'' $\omega_p\tau_0 > 10$,
 so that the parametric gain in the hyperbolic time crystal (see Fig. \ref{fig:gain}(b)) can reach and even exceed the materail 
 absorption -- making it possible to bring the dream of a lossless hyperbolic material to the reality of practical engineering.

\section{Practical Considerations}

With the present stage of development in the field of low-loss optical materials with free charge carriers, 
highly doped semiconductors and transparent conducting oxides (TCOs) can support the proposed
approach of the carrier drift modulation, suitable respectively for mid- and near-infrared wavelength range.
With $\omega_p \tau_0 \sim 5$ for TCOs \cite{ITO} and $\omega_p \tau_0 > 10$ for doped semiconductors \cite{nmat},
the parametric gain in the time hypercrystal (see Fig. \ref{fig:gain}(b)) can overcome the material absorption in both of 
these platforms.

Furthermore, with the signal in the mid- or near-infrared frequency bands, using a few femtosecond pump pulse 
will satisfy the drift modulation requirement of $\tau_M \ll 2 \pi/\omega$.

However, to reach the hyperbolic regime, the impulse $p_1$ transferred to the free electrons from the pump pulse
(see the inequality (\ref{eq:h-band})) must reach the values when
\begin{eqnarray}
\frac{\alpha_* p_1^2}{m_0} \sim 0.1,
\end{eqnarray}
so that the peak amplitude of the pump pulse
\begin{eqnarray}
E_M & \sim & \frac{p_1}{e \tau_M} \sim \sqrt{\frac{0.1 m_0}{\alpha_* e^2 \tau_M^2}},
\end{eqnarray}  
For Ga$_{0.47}$In$_{0.53}$As we find $m_0 =  0.41\, m_e$, where $m_e$ is the electron mass, and $\alpha_* \simeq
1.35 \ {\rm eV}^{-1}$ \cite{Capasso1985}, so that for $\tau_M \sim 3$ fsec we find $E_M \sim 500$ kV/cm, which is 
several orders of magnitude below the  peak amplitudes of $ \sim 100$ MV/cm of  commercial single-cycle IR 
sources \cite{Lenke2024}.

The drift modulation approach introduced in the present work, therefore allows to induce  a loss-free hyperbolic time 
crystal in an existing and well-established semiconductor material platform, using only commercially available 
sources.

\section{Conclusions}

In summary, we have introduced Carrier Drift Modulation, a novel mechanism for creating temporal boundaries and enabling the formation of photonic time crystals. This approach paves the way toward hyperbolic temporal metamaterials and hyperbolic time crystals, establishing a new paradigm in time‑domain photonics. Importantly, we show that the very process responsible for forming the hyperbolic time crystal can simultaneously compensate intrinsic material losses in the supporting medium. Remarkably, such lossless hyperbolic time crystals can be realized using existing materials and readily available light sources, underscoring both the practicality and transformative potential of this concept.

\section{Acknowledgements}

B.S.  is grateful to Ohad Segal for useful discussions.
\appendix

\section{The Wave Equation \label{sec:A-WE}}

With the linear-response distribution function $g$ given by Eqn. (\ref{eq:g}),
for the corresponding current density we obtain
\begin{eqnarray}
{\bf j} & = & \frac{n_0 e^2 }{\tau_0} 
 \int_{-\infty}^t  {dt'} \int_{-\infty}^{t'} {dt''}  \,   \,  \exp\left({ -\frac{t - t''}{\tau_0}}\right) 
\nonumber \\
& \times & \frac{1}{{\bf m}\left[ {\bf p}_M\left({\bf r}, t\right)  - {\bf p}_M\left({\bf r},t''\right)  \right]} 
\cdot {\bf E}\left({\bf r}, t'\right), \label{eq:j1}
\end{eqnarray}
where the modulated  effective mass tensor
\begin{eqnarray}
\frac{1}{m_{\alpha\beta}\left[{\bf q}\right]} & = & 
\biggl< \frac{\partial^2 \varepsilon_{\bf p+ q}}{\partial p_\alpha \partial p_\beta}  \biggr>_{\bf p}  
\end{eqnarray}
with the phase space average defined as 
\begin{eqnarray}
  \bigl< \,  F\left[{\bf p +  q}\right] \, \bigr>_{\bf p}  
  & \equiv & 
  \frac{\int d{\bf p} \, F\left({\bf p + q}\right) f_0\left({\bf p}\right)}{\int d{\bf p} \,  f_0\left({\bf p}\right)}.
 \end{eqnarray} 
Note that for a parabolic band 
\begin{eqnarray}
  \biggl< 
  \frac{\partial^2  \varepsilon_{{\bf p+q}} }{\partial p_\alpha \partial p_\beta} \biggr>_{\bf p}  & = & 
  \biggl< \frac{\partial^2  \varepsilon_{{\bf p}} }{\partial p_\alpha \partial p_\beta}  \biggr>_{\bf p} \equiv 
  \frac{\delta_{\alpha\beta}}{m_*}
\end{eqnarray}
is defined by the average effective mass $m_*$, so that 
the current density (\ref{eq:j1}) does not depend on the pump, and  the modulation has no effect on the free
carrier response to the probe field.
 
 Introducing the modulation kernel 
 \begin{eqnarray}
 {\bf M}\left( t, t'\right) & \equiv & \omega_p^2  \int_{-\infty}^{t'} \frac{dt''}{\tau_0}  \, \exp\left({ \frac{ t''}{\tau_0}}\right) 
\nonumber \\
& \times & \left( 1  - \frac{m_*}{{\bf m}\left[ {\bf p}_M\left(t\right)  - {\bf p}_M\left(t''\right)  \right]}  \right), 
 \end{eqnarray} 
where the plasma frequency
\begin{eqnarray}
\omega_p^2 & = & \frac{4\pi n_0 e^2}{m_* \epsilon_\infty},
\end{eqnarray}
and  $\epsilon_\infty$  is the ``background'' dielectric permittivity from the crystal lattice, we obtain
\begin{eqnarray}
{\bf j} & = & \frac{\epsilon_\infty  \, \omega_p^2}{4 \pi} \,  
 \int_{-\infty}^t  {dt'}  \, e^{{ -\frac{t - t'}{\tau_0}}} \, 
 {\bf E}\left({\bf r}, t'\right)
\nonumber \\
& - & \frac{\epsilon_\infty  }{4 \pi}   \, e^{{ -\frac{t}{\tau_0}}}
\int_{-\infty}^t  {dt'}  \, {\bf M}\left(t, t'\right)  
\cdot  {\bf E}\left({\bf r}, t'\right), \label{eq:j2A}
\end{eqnarray}

Note that at the time scale  when the electron relaxation can be neglected, the modulation kernel reduces to
\begin{eqnarray}
{\bf M}\left(t, t'\right) & = & \omega_p^2 \left( 1  - \frac{m_*}{{\bf m}\left[ {\bf p}_M\left( t\right) \right]}  \right)
\end{eqnarray}

In a non-magnetic medium, the evolution of the probe electric field ${\bf E}\left({\bf r}, t\right)$ is defined by the wave equation
\begin{eqnarray}
\epsilon_\infty \frac{\partial^2 {\bf E}}{\partial t^2} + c^2 \, {\rm curl}\, {\rm curl} \, {\bf E}
+ 4 \pi \frac{\partial {\bf j}}{\partial t} 
 & = & 0.
 \label{eq:we1}
\end{eqnarray}
From Eqn. (\ref{eq:j2A}),
\begin{eqnarray}
& & \left(\frac{\partial}{\partial t}  +   \frac{1}{\tau_0} \right) {\bf j}\left({\bf r}, t\right)
=   \frac{\epsilon_\infty  \, \omega_p^2}{4 \pi}  \, {\bf E}\left({\bf r}, t\right) - \frac{\epsilon_\infty }{4 \pi}   \,  e^{ -\frac{t}{\tau_0}}   \nonumber \\
& \times  &  \, 
\left[  {\bf M}\left(t, t\right)\, {\bf E}\left({\bf r}, t\right) +  \int_{-\infty}^t  {dt'}  \, \frac{\partial {\bf M}\left(t, t'\right)}{\partial t}  
\cdot  {\bf E}\left({\bf r}, t'\right) \right], \ \ \ \ \ \ \ \label{eq:j3A}
\end{eqnarray}
so that we can express (\ref{eq:we1}) as the Wave Equation
\begin{eqnarray}
& & \left(\frac{\partial}{\partial t}    +    \frac{1}{\tau_0} \right) \left(   \frac{\partial^2 {\bf E}}{\partial t^2} + 
\frac{c^2}{\epsilon_\infty}  \, {\rm curl}\, {\rm curl} \, {\bf E}  \right)
+  \omega_p^2 \ \frac{\partial{\bf E} }{\partial t} \nonumber \\
& = &  \frac{\partial }{\partial t}    \left[  {\bf M}_0 \cdot {\bf E}
+  e^{ -\frac{t}{\tau_0}} \int_{-\infty}^t  {dt'}  \, \, \frac{\partial {\bf M}}{\partial t}  
\cdot  {\bf E}\left(t'\right) \right], \ \ \ \ \ \ \ \label{eq:we2A}
\end{eqnarray}
where 
\begin{eqnarray}
{\bf M}_0\left( t \right) \equiv {\bf M}\left( t, t \right) \,  e^{ -\frac{t}{\tau_0}}.  
\label{eq:M0A}
\end{eqnarray}

\section{Boundary Conditions at a Time Interface \label{sec:A-BC}}

The original wave equation (\ref{eq:we1}) that is a direct consequence of the full set of Maxwell's Equations, implies the continuity
of the electric field ${\bf E}\left(t\right)$ and the field 
\begin{eqnarray}
{\bf G} & \equiv & \frac{\partial {\bf E}}{\partial t}  - \frac{4\pi}{\epsilon_\infty} {\bf j}. 
\label{eq:AG}
\end{eqnarray}
Substituting (\ref{eq:j2A}) into (\ref{eq:AG}),  we find
\begin{eqnarray}
{\bf G}  & = & {\bf F} +  \omega_p^2  \int_{-\infty}^{t} {dt'}  \, e^{- \frac{ t-t'}{\tau_0}} \, {\bf E}\left(t'\right),
\label{eq:AGF}
\end{eqnarray}
where the auxiliary field
\begin{eqnarray}
{\bf F}& \equiv & \frac{\partial {\bf E}}{\partial t} - e^{{ -\frac{t}{\tau_0}}}
\int_{-\infty}^t  {dt'}  \, {\bf M}\left(t, t'\right)  
\cdot  {\bf E}\left({\bf r}, t'\right). \label{eq:AF}
\end{eqnarray}
Therefore, together with  the already established continuity of the electric field ${\bf E}$, Eqn. (\ref{eq:AF}) implies the continuity 
of the auxiliary ``time displacement field'' ${\bf F}$.

Then the wave equation (\ref{eq:we2}) can be expressed as
\begin{eqnarray}
 \frac{\partial}{\partial t}\left[ \frac{\partial{\bf F}}{\partial t}   + \frac{\bf F}{\tau_0}
  \right] & + & \frac{\partial}{\partial t}\left[ \frac{c^2}{\epsilon_\infty}  \, {\rm curl}\, {\rm curl} \, {\bf E} +  \omega_p^2 \, {\bf E} \right]\nonumber \\
&  + &  \frac{1}{\tau_0} \, {\rm curl}\, {\rm curl} \, {\bf E} 
  =  0, 
  \label{eq:we4}
\end{eqnarray}
which implies the continuity of the derivative $\partial {\bf F}/\partial t$. 

Therefore, the complete set of three boundary conditions
at a time interface, corresponds to the continuity of the electric field ${\bf E}$, the time displacement field ${\bf F}$ and the first (time)
derivative of the time displacement field $\partial {\bf F}/\partial t$. 

\section{Wave Equation in the Lossless Limit \label{sec:A-LL-we}}

In the nearly-lossless limit $\omega \tau_0 \gg 1$, we can neglect the scattering term in the kinetic equations (\ref{eq:KE1}) and (\ref{eq:g}), which yields
\begin{eqnarray}
f_M\left({\bf p}, t\right) & = & f_0\left({\bf p} - {\bf p}_D\left(t\right) \right), 
\end{eqnarray}
and
\begin{eqnarray}
g & = & \frac{\partial f_0\left( {\bf p} - {\bf p}_M\left(t\right)  \right)}{\partial {\bf p} }
\int_{-\infty}^t dt' \, e{\bf E}\left( t\right), 
\end{eqnarray}
so that the current density due to the signal field ${\bf E}$  is given by
\begin{eqnarray}
{\bf j}\left({\bf r}, t\right)  & = & \frac{\epsilon_\infty}{4\pi} \, \omega_p^2 \,  \frac{m_*}{{\bf m}_M} \, 
\int_{- \infty}^t dt' \, e {\bf E}\left({\bf r}, t'\right),
\label{eq:Aj-LL}
\end{eqnarray}
with the time-dependent effective mass  defined as
\begin{eqnarray}
\left[ \frac{m_*}{{\bf m}_M\left(t\right)} \right]_{\alpha\beta}  & = & 
\frac{\bigl< 
  \frac{\partial^2  \varepsilon_{{\bf p}+{\bf p}_M\left(t\right)  }}{\partial p_\alpha \partial p_\beta } \bigr>_{\bf p}   }{\bigl< 
  \frac{\partial^2  \varepsilon_{{\bf p} }}{\partial p^2 } \bigr>_{\bf p}  }.
\end{eqnarray}
Substituting (\ref{eq:Aj-LL}) into (\ref{eq:we1}), we obtain the wave equation (\ref{eq:weLL0}).

\section{Electronic Band Non-parabolicity \label{sec:A}}

An accurate description for a non-parabolic band \cite{Balkaitski1968} in an optoelectronic material can be obtained from the celebrated Kane model
\cite{Kane1957} that
originates from a perturbation approach to the solution of
the Schr\"odinger equation of a single electron in a crystal potential including the spin–orbit interaction:
\begin{eqnarray}
\varepsilon_{\bf p} \left(1 + \alpha_* \varepsilon_{\bf p}\right) & = & \frac{p^2}{2 m_0},
\label{eq:Kane}
\end{eqnarray}
where the (temperature-dependent) parameter $\alpha$ is on the order of (and in some cases such e.g., bismuth and its allows, equal to) the inverse bandgap energy $\alpha_* \sim E_g^{-1}$.
While highly accurate for the 
direct bandgap III–V semiconductors of cubic symmetry and their alloys, the Kane model (\ref{eq:Kane}) can also be used 
at the quantitative level for many other materials \cite{Zawaldski1973,AMbook,Green1990,Masut2022}.

With the Kane's expression for $\varepsilon_{\bf p}$, we obtain
\begin{eqnarray}
\frac{\partial^2 \varepsilon}{\partial p_\alpha  \partial p_\beta}
& = & 
\frac{1}{m_0} \left[ \frac{\delta_{\alpha\beta} }{1 + 2 \alpha_* \varepsilon_{\bf p}} - 
\frac{2 \, \alpha_* p_\alpha p_\beta}{m_0 \left( 1 + 2 \alpha_* \varepsilon_{\bf p} \right)^3} \right]. \ \ \ \  
\end{eqnarray} 
so that  for the induced anisotropy in the dielectric permittivity (\ref{eq:eps}) we find 
\begin{eqnarray}
 \frac{\Omega_\parallel^2 - \Omega_\perp^2  }{\omega_p^2}& = &  2 \, \frac{ \alpha_*}{m_0}  \sum_{t_n < t} e^{- \frac{t - t_n }{\tau_0}}  \Delta p_n^2
 \left(1 - \Delta_n\right),  
\end{eqnarray}
where 
\begin{eqnarray}
\Delta_n & = &    \frac{e^2}{2 \pi^2 \hbar^3 \epsilon_\infty  \omega_p^2} \int d{\bf p} \, \frac{ f_0\left({\bf p - \Delta p}_n\right) -f_0\left({\bf p}\right)  }{\Delta p_n^2}  \nonumber \\
& \times & \left(p^2 -  \frac{3 \left({\bf p}\cdot{\bf \Delta p}_n\right)^2}{\Delta p_n^2} \right) \left(1 - \frac{1}{\left( 1 + 2 \alpha_* \varepsilon_{\bf p} \right)^3}   \right) \ \ \ \ \\
& = & {\cal O}\left(\alpha_* \varepsilon_F\right) = {\cal O}\left(\frac{\varepsilon_F}{E_g}\right),
\end{eqnarray}
and the subscripts correspond to the directions parallel ($\parallel$) and perpendicular ($\perp$) to the {\it pump} field
${\bf E}_M$ modulating the material. 

In the limit of moderate carrier drift modulation, $\Delta p_n < p_F$, we find
\begin{eqnarray}
\Omega_\parallel^{\ 2}  & \simeq & {3} \frac{\alpha_* \omega_p^2}{m_0} \, 
 \sum_{t_n < t}   \Delta p_n^2 \, e^{- \frac{t - t_n }{\tau_0}},  \\ 
\label{eq:WprlA}
\Omega_\perp^{\ 2} & \simeq & \frac{\alpha_* \omega_p^2}{m_0} \, 
 \sum_{t_n < t}   \Delta p_n^2 \, e^{- \frac{t - t_n }{\tau_0}}. 
\label{eq:WprpA}
\end{eqnarray}

\section{Time-Reflection and Time-Transmission Coefficients -- general treatment \label{sec:A-TRT}}

Close to the time interface at $t=0$ (i.e. for $t \ll \tau_0$) with a single monochromatic incident wave
(\ref{eq:Ei}),   the auxiliary field ${\bf F}$ and its derivative are given by
\begin{eqnarray}
{\bf F} & = & \frac{\partial {\bf E}}{\partial t} - \Omega_\perp^2  \theta\left(t\right) \int_{-\infty}^0 dt' \,  
e^{ t'/{\tau_0}}  \, {\bf E}\left(t'\right) \nonumber \\
& = &  \frac{\partial {\bf E}}{\partial t} - \frac{ \Omega_\perp^2 \tau_0  }{1 - i \omega \tau_0} \theta\left(t\right) {\bf E},
\label{eq:F} \\
\frac{\partial {\bf F}}{\partial t}  & = & \frac{\partial^2 {\bf E}}{\partial t^2} -
\Omega_\perp^2  \theta\left(t\right)   \left( {\bf E} - \int_{-\infty}^0 \frac{dt'}{\tau_0}  \,  
e^{ t'/{\tau_0}}  \, {\bf E}\left(t'\right)  \right) \nonumber \\
& = &  \frac{\partial^2 {\bf E}}{\partial t^2} - \frac{ \Omega_\perp^2   }{1 +\frac{ i}{ \omega \tau_0}} \theta\left(t\right) {\bf E}.
\label{eq:dF} 
\end{eqnarray}

Using the continuity of the fields ${\bf E}$, ${\bf F}$ and $\partial {\bf F}/\partial t$ (see Eqn. (\ref{eq:F1})  in Section \ref{sec:we})  at the temporal interface $t=0$,
from Eqns.   (\ref{eq:Ei}),  (\ref{eq:F}), (\ref{eq:dF}) and (\ref{eq:Es2})  we obtain
\begin{eqnarray}
& &{ \cal T}   +  {\cal R} + { \cal S}  =  1, \\
& & \left(\omega' - \frac{ i \gamma}{\tau_0}\right) { \cal T} + \left(-\omega' -  \frac{i \gamma}{\tau_0}\right) {\cal R} 
 - i   \frac{1 - 2 \gamma}{\tau_0} { \cal S}  
\nonumber \\
& & \ =  
\omega \left( 1 - \frac{\Omega_\perp^2}{\omega\left(\omega + i/\tau_0\right)}\right), \ \ \ \\
& &  \left(\omega' - \frac{ i \gamma}{\tau_0}\right)^2 { \cal T}  + \left(\omega' +  \frac{i \gamma}{\tau_0}\right)^2 {\cal R}
-   \frac{\left(1 - 2 \gamma\right)^2 }{\tau_0^2} 
\, { \cal S}  \nonumber  \\
&  & \  =  
 \omega^2 \left( 1 - \frac{\Omega_\perp^2}{\omega\left(\omega + i/\tau_0\right)}\right) 
\end{eqnarray}
or equivalently
\begin{eqnarray}
&  & {\cal W}_3 \left[ \omega' - \frac{i\gamma}{\tau_0}, - \omega' - \frac{i\gamma}{\tau_0} ,  
- i \frac{\left(1 - 2 \gamma\right) }{\tau_0}  \right]
\left(
\begin{array}{c}
{\cal T} \\
{\cal R} \\
{\cal S} 
\end{array}
\right) \nonumber \\
& \ & \ \ =    
\left(
\begin{array}{c}
1 \\
\eta_\omega \, \omega  \\
\eta_\omega \, \omega^2  
\end{array}
\right),
\end{eqnarray}
where 
\begin{eqnarray}
\eta_\omega & \equiv & 1 - \frac{\Omega_\perp^2}{\omega\left(\omega + i/\tau_0\right)},
\end{eqnarray}
and ${\cal W}_3$ is the 3rd order Vandermonde matrix \cite{Macon1958} 
\begin{eqnarray}
{\cal W}_3\left(x_0, x_1, x_2\right) \equiv \left[
\begin{array}{ccc}
1 & 1 & 1 \\
x_0 & x_1 & x_2 \\
x_0^2 & x_1^2 & x_2^2
\end{array}
\right]
\end{eqnarray}
with the determinant 
\begin{eqnarray}
{\rm det} \, {\cal W}_3 & = & (x_0 - x_1) (x_0 - x_2) (x_1 - x_2), 
\end{eqnarray}
and the inverse
\begin{eqnarray}
 {\cal W}_3^{-1} & = & 
\frac{1}{{\rm det} \ {\cal W}_3 }  \nonumber \\& \times & 
\left[
\begin{array}{ccc}
x_1 x_2 \left( x_2 - x_1\right) &  x_1^2 - x_2^2 & x_2 - x_1  \\
x_0 x_2 \left( x_0 - x_2\right) &  x_2^2 - x_0^2 & x_0 - x_2 \\
x_0 x_1 \left( x_1 - x_0\right) & x_0^2 - x_1^2 & x_1 - x_0 
\end{array}
\right]. 
\end{eqnarray}
We therefore obtain
\begin{eqnarray}
{\cal R} & = & \frac{\eta_\omega \omega ( \omega - \omega' + i \frac{1 -  \gamma}{\tau_0} ) 
- i \frac{1 - 2 \gamma}{\tau_0} 
(\omega' - i \frac{ \gamma}{\tau_0} ) }{2 \, \omega' \, \left( \omega' - i\frac{1 - 3 \gamma}{\tau_0} \right) }, \\
{\cal T} & = & \frac{\eta_\omega \omega ( \omega + \omega' + i \frac{1 -  \gamma}{\tau_0} ) 
+i \frac{1 - 2 \gamma}{\tau_0} 
(\omega' + i \frac{ \gamma}{\tau_0} ) }{2 \, \omega' \, \left( \omega' + i\frac{1 - 3 \gamma}{\tau_0} \right) },
 \\
{\cal S} & = & \frac{ {\omega'}^2  - \eta_\omega \omega \left( \omega + 2 i  \gamma/\tau_0\right)  + \gamma^2 / \tau_0^2}{{\omega'} ^2+{\left(1 - 3 \gamma\right)^2}/{\tau_0^2} },
\end{eqnarray}

\section{Time-Reflection and Time-Transmission Coefficients -- lossless limit \label{sec:A-TRT-LL}}

In the lossless limit $\omega \tau_0 \gg 1$ the wave equation (\ref{eq:weLL0}) is of the second order in time, so that the time-evanescent field is decoupled from the propagating waves and is not excited at a temporal interface. When the modulation pulse applied at $t=0$, for 
the electric field we therefore obtain
\begin{eqnarray}
{\bf E}\left( {\bf r}, t\right) & = & {\bf E}_0 \, e^{i {\bf k}\cdot{\bf r} } \,
\left\{
\begin{array}{cc}
e^{- i \omega t } , & t < 0 \\
{\cal T} e^{- i \omega' t } + {\cal R} e^{ i \omega' t } , & t > 0.
\end{array} 
\right.
\end{eqnarray}
where  (see also Eqns. (\ref{eq:w-initial}), (\ref{eq:wprime}))
\begin{eqnarray}
\omega & = & \sqrt{\omega_p^2 + \frac{k^2 c^2}{\epsilon_\infty}}, \label{eq:Aw} \\
\omega' & = &  \sqrt{\frac{m_*}{m_M} \, \omega_p^2  + \frac{k^2 c^2}{\epsilon_\infty}}.  \label{eq:Awp} \
\end{eqnarray}
and, in the coordinates of Fig. \ref{fig:waveguide},
\begin{eqnarray}
m_M & = & \left[ {\bf m}_M\left(t>0\right) \right]_{zz} \equiv  
{\biggl< 
  \frac{\partial^2  \varepsilon_{{\bf p}+{\bf p}_M  }}{\partial p_z^2 } \biggr>_{\bf p}^{-1} }. 
\end{eqnarray}
Then
\begin{eqnarray}
\int_{-\infty}^0 dt' \, {\bf E}\left(t'\right) & = & - \ \frac{{\bf E}_0  \, e^{i {\bf k}\cdot{\bf r} } }{i\omega}, 
\end{eqnarray} 
and the temporal displacement at the time interface $t = 0$
\begin{eqnarray}
{\bf F} & = & 
 {\bf E}_0 \, e^{i {\bf k}\cdot{\bf r} } \,
\left\{
\begin{array}{cc}
- i \, \omega \, \left[ 1 - \frac{   \omega_p^2}{\omega^2} \right], & t = - \, 0, \\
- i \, \omega' \, \left[ {\cal T} - {\cal R}  -  \frac{m_*}{m_M} \, \frac{ \omega_p^2}{\omega'^2}\right], & t = + \, 0.
\end{array} 
\right.
\ \ \ \ \ \ 
\end{eqnarray}
Using the continuity of the electric field ${\bf E}$ and the time displacement ${\bf F}$ at $t=0$,  
we obtain
\begin{eqnarray}
{\cal T} & = & \frac{1}{2} +   \frac{\omega^2 -  \omega_p^2 \left(  1 -  \frac{m_*}{m_M} \right) }{2 \omega \omega'}, \label{eq:AT} \\
 {\cal R} & = & \frac{1}{2}  - \frac{\omega^2 -  \omega_p^2 \left(  1 -  \frac{m_*}{m_M} \right)}{2 \omega  \omega'}. \label{eq:AR}
\end{eqnarray}
From Eqns. (\ref{eq:Aw}),(\ref{eq:Awp})
\begin{eqnarray}
{\omega'}^2 & = & \omega^2 -  \omega_p^2 \left(  1 -  \frac{m_*}{m_M} \right),
\end{eqnarray}
which reduces Eqns. (\ref{eq:AT}),(\ref{eq:AR}) to 
\begin{eqnarray}
{\cal T} & = &  \frac{\omega + \omega'}{2 \omega}, \\
 {\cal R} & = &  \frac{\omega - \omega'}{2 \omega}. 
\end{eqnarray}

\section{Wave Equation in the Quasistatic Limit \label{sec:A-WE-QS}}

\noindent
From the Gauss Law
\begin{eqnarray}
\epsilon_\infty \, {\rm div}{\bf E} & = & 4 \pi \rho
\end{eqnarray}
and the  conservation of free charge
\begin{eqnarray}
\frac{\partial\rho}{\partial t} + {\rm div}{\bf j} & = & 0,
\end{eqnarray}
we obtain
\begin{eqnarray}
\epsilon_\infty \frac{\partial}{\partial t} \nabla^2 \phi & = &  4 \pi \, \nabla\cdot{\bf j}.
\label{eq:phiA}
\end{eqnarray}
Introducing the Fourier representation of the scalar potential
\begin{eqnarray}
\phi\left({\bf r}, t\right) & = &  \int \phi_{\bf k}\left(t\right)  \, e^{i {\bf k}\cdot{\bf r}},
\label{eq:phikA}
\end{eqnarray}
and substituting Eqn. (\ref{eq:j2A}) into (\ref{eq:phiA}), we obtain
\begin{eqnarray}
 e^{  \frac{t }{\tau_0}}  \  \frac{d\phi_{\bf k} }{d t}  &  + &  
\omega_p^2  \, \int_{-\infty}^t dt' \, \phi_{\bf k}\left(t'\right) e^{  \frac{ t'}{\tau_0}}  \nonumber \\
& = &  \int_{-\infty}^t dt' \, {\cal M}_{\bf k}\left(t,t'\right) \,  \phi_{\bf k}\left( t'\right) , 
\label{eq:phi1A}
\end{eqnarray}
where the (scalar) modulation kernel ${\cal M}_{\bf k}\left(t\right)$ is given by
\begin{eqnarray}
{\cal M}_{\bf k}\left(t, t'\right) & = & \frac{{\bf k}\cdot{\bf M}\left(t, t' \right)\cdot{\bf k}}{k^2}. 
\end{eqnarray}
For the wave equation in the quasistatic limit we therefore obtain
\begin{eqnarray}
 \frac{d^2\phi_{\bf k} }{d t^2} & + & \frac{1}{\tau_0} \frac{d\phi_{\bf k} }{d t}  +   
 \left( \omega_p^2  - {\cal M}_{\bf k}^{(o)}\left(t\right) \right) \phi_{\bf k} 
 \nonumber \\
 & = & e^{ - \frac{ t}{\tau_0}} \,  \int_{-\infty}^t dt' \, \frac{\partial {\cal M}_{\bf k}}{\partial t}  \,  \phi_{\bf k}\left( t'\right), 
\label{eq:phi2A}
\end{eqnarray}
where
\begin{eqnarray}
{\cal M}_{\bf k}^{(0)}\left(t\right)  & \equiv & e^{ - \frac{ t}{\tau_0}} \, {\cal M}_{\bf k}\left(t,t\right).  
\end{eqnarray}


\begin{thebibliography}{9}
\bibitem{explain}
E.g., for heavily doped semiconductors used in near to mid infrared (IR) range such as GaAs \cite{nmat} , $c \, \tau_0 \sim  100 \ \mu{\rm m}$, while in transparent conducting oxides (wavelengths in visible to near IR) such as ITO \cite{Julia}, $c \, \tau_0 \sim  10 \ \mu{\rm m}$. In both cases the length $c \tau_0$ exceeds the corresponding values of $1/k \sim \lambda_0$, (where $\lambda_0$ is the free-space signal wavelength) by more than an order of magnitude.
\end{thebibliography}
\end{document}